\documentclass[a4paper,11pt]{article}
\pdfoutput=1
\usepackage{jheppub}
\usepackage[T1]{fontenc}
\usepackage[utf8]{inputenc}
\usepackage{newunicodechar}
\newunicodechar{，}{,}
\usepackage{graphicx}
\usepackage[dvipsnames]{xcolor}

\usepackage{theorem}
\allowdisplaybreaks     
\usepackage{subfigure}
\usepackage{booktabs}
\usepackage[percent]{overpic}
\usepackage{amssymb}
\usepackage{amsfonts}
\usepackage{amsmath}
\usepackage{bm}
\usepackage{hyperref} 
\usepackage{enumitem}
\usepackage{cleveref}
\numberwithin{equation}{section}
\renewcommand{\arraystretch}{1.5}

\usepackage{tikz}
\usetikzlibrary{shapes.geometric, arrows, positioning, calc}

\tikzstyle{startstop} = [rectangle, rounded corners, minimum width=3cm, minimum height=1cm,text centered, draw=black, fill=red!30]
\tikzstyle{io} = [trapezium, trapezium left angle=70, trapezium right angle=110, minimum width=3cm, minimum height=1cm, text centered, draw=black, fill=blue!30]
\tikzstyle{process} = [rectangle, minimum width=3cm, minimum height=1cm, text centered, draw=black, fill=orange!30]
\tikzstyle{decision} = [diamond, minimum width=3cm, minimum height=1cm, text centered, draw=black, fill=green!30]
\tikzstyle{arrow} = [thick,->,>=stealth]

{\theorembodyfont{\rmfamily}}

\newtheorem{theorem}{Theorem}

\title{\boldmath Probing bulk geometry via pole skipping: from static to rotating spacetimes}   
\author[a]{Cheng Ran,}
\author[a,c]{Zhenkang Lu,}
\author[a,b]{and Shao-Feng Wu}

\affiliation[a]{Department of Physics, Shanghai University, Shanghai, 200444, China}

\affiliation[b]{Center for Gravitation and Cosmology, Yangzhou University, Yangzhou 225009, China}

\affiliation[c]{School of Science and Engineering, The Chinese University of Hong Kong (Shenzhen), Longgang, Shenzhen, Guangdong, 518172, China}

\emailAdd{r\_cheng@shu.edu.cn}
\emailAdd{zhenkanglu9@gmail.com}
\emailAdd{sfwu@shu.edu.cn}
\abstract{
We investigate an analytical framework for reconstructing bulk geometries from pole-skipping data. Previously, this method enabled the recursive recovery of near-horizon metric derivatives in static, planar-symmetric black holes. Building on this framework, we systematically extend it to more intricate geometries, specifically static topological black holes and rotating black holes. For three-dimensional rotating black holes, we demonstrate
that the metric can be fully reconstructed from boundary pole-skipping data. For
four-dimensional rotating spacetimes admitting a separable coordinate system
(such as the Kerr family), standard near-horizon pole-skipping successfully
reconstructs the purely radial metric functions. To recover the remaining angular
metric functions, we introduce a mathematical counterpart termed ``angular
pole-skipping,'' defined via a near-axis analysis. Although its precise holographic
dictionary remains an open question, this bulk-side formalism completes the
geometric reconstruction algorithm. Furthermore, we demonstrate that the vacuum Einstein equations can be recast as a set of algebraic equations governing the pole-skipping data, and that the Null Energy Condition imposes algebraic inequalities on this boundary data.
Finally, we establish general polynomial constraints dictated by the
overdetermined nature of the metric reconstruction, highlighting the highly redundant
encoding of bulk geometry in boundary data.}

\begin{document}
\maketitle
\flushbottom

\section{Introduction}
\label{sec:intro}
Holographic duality \cite{Witten:1998qj, Gubser:1998bc, Maldacena:1997re} relates the dynamics of black holes in the bulk to thermal many-body physics on the boundary. In this framework, the high-energy gravitational scattering near the horizon, described by gravitational shockwaves, provides a bulk mechanism for scrambling and quantum chaos in the dual field theory~\cite{Shenker:2013pqa, Shenker:2013yza, Roberts:2014isa, Shenker:2014cwa}. Quantitatively, this chaotic behavior is diagnosed by the exponential growth of out-of-time-order correlators (OTOCs). The associated Lyapunov exponent $\lambda_L$ is subject to the Maldacena-Shenker-Stanford (MSS) bound~\cite{Maldacena:2015waa}, namely $\lambda_L \le 2\pi T$ (in natural units). Remarkably, this fundamental limit is strictly saturated by large-$N$, strongly coupled theories that admit an Einstein gravity dual.

Another manifestation of quantum chaos, known as pole skipping~\cite{Grozdanov:2017ajz, Blake:2017ris, Blake:2018leo, Natsuume:2019xcy}, has been identified in large-$N$ systems that saturate the maximal Lyapunov exponent. This phenomenon refers to the fact that at special points in the complex frequency and momentum plane, the retarded Green's function of the energy density operator becomes ill-defined. This occurs due to the intersection of lines of poles and zeros, resulting in an indeterminate $0/0$ form. While the leading pole-skipping point in the upper-half plane allows for the direct extraction of the Lyapunov exponent and butterfly velocity, an infinite tower of such points emerges at negative imaginary Matsubara frequencies. Essentially, this phenomenon reveals a universal feature of finite-temperature systems with gravitational duals: the near-horizon expansion of bulk perturbation equations admits an extra free parameter at these special frequencies, consequently failing to uniquely determine a regular solution~\cite{Blake:2019otz}.\footnote{ Following this realization, pole-skipping has been extensively explored across diverse gravitational models and various perturbation sectors~\cite{Blake:2021hjj, Natsuume:2019sfp, Ahn:2019rnq, Liu:2020yaf, Ramirez:2020qer, Grozdanov:2018kkt, Grozdanov:2019uhi, Natsuume:2019vcv, Wu:2019esr, Yuan:2020fvv, Yuan:2021ets, Baishya:2023mgz, Yuan:2023tft, Yuan:2024utc, Ahn:2025exp, Ceplak:2019ymw, Sil:2020jhr, Ahn:2020bks, Abbasi:2020ykq, Grozdanov:2020koi, Ahn:2020baf, Natsuume:2020snz, Abbasi:2019rhy, Kim:2020url, Abbasi:2020xli, Ceplak:2021efc, Jeong:2021zhz, Kim:2021xdz, Wang:2022mcq, Amano:2022mlu, Natsuume:2023hsz, Ning:2023ggs, Jeong:2023rck, Ferreira:2025iqe, Yuan:2025ivz, Asplund:2025nkw, Grozdanov:2025ulc, Gao:2025ohk, Lyu:2025kep, Davison:2025xdj, Natsuume:2025nkx}.}

Framing this near-horizon analysis as an inverse problem, we recently developed a novel scheme that reconstructs the bulk geometry—including the black hole interior—from boundary pole-skipping data in a fully analytical manner~\cite{Lu:2025jgk, Lu:2025pal}. The core steps are summarized as follows:
\begin{itemize}
\item Perform a near-horizon expansion of the Klein-Gordon equation governing the decoupled perturbations and evaluate it at the special Matsubara frequencies where the system admits an extra free parameter. Cast the resulting truncated linear system into matrix form and require the determinant of the associated coefficient matrix to vanish, yielding a characteristic polynomial in the pole-skipping momentum.
\item Establish that at each order $n$, this polynomial depends linearly on the $n$th-order metric derivatives at the horizon.
\item By applying Vieta's formulas, this polynomial is translated into a set of linear equations for the $n$th-order metric derivatives, which are then solved order by order. This allows the metric derivatives to be expressed analytically to an arbitrary order in terms of the boundary pole-skipping data.
\end{itemize}

Beyond our reconstruction scheme based on pole skipping, there are numerous established approaches to recovering the classical bulk geometry and background matter fields, as well as local bulk operators. Traditionally, these methods rely on extracting data from diverse boundary observables, including boundary $n$-point functions and related features \cite{deHaro:2000vlm, Hubeny:2006yu, Hammersley:2006cp, Bilson:2008ab, Qi:2013caa, Engelhardt:2016wgb, Engelhardt:2016crc, Fan:2023bsz, Fan:2025fxt, Hashimoto:2022aso, Caron-Huot:2022lff, Nebabu:2023iox, Nebabu:2026lmn}, subregion entropies \cite{Bilson:2010ff, Nozaki:2012zj, Czech:2015qta, Hammersley:2007ab, You:2017guh, Hubeny:2012ry, Roy:2018ehv, Myers:2014jia, Balasubramanian:2013lsa, Czech:2014ppa, Jokela:2023rba, Jokela:2020auu}, complexity \cite{Hashimoto:2021umd, Xu:2023eof}, Wilson loops \cite{Hashimoto:2020mrx}, modular Hamiltonians of boundary subregions \cite{Kabat:2018smf}, and non-local boundary operators~\cite{Hamilton:2005ju, Dong:2016eik}. Furthermore, to construct and decode this complex boundary-to-bulk mapping, powerful theoretical models and computational algorithms have been extensively employed, most notably tensor networks~\cite{Swingle:2009bg, Pastawski:2015qua, Cao:2020uvb, Vidal:2008zz} and machine learning~\cite{Hashimoto:2018ftp, Hashimoto:2018bnb, Yan:2020wcd, Akutagawa:2020yeo, Hashimoto:2020jug, Hashimoto:2021ihd, Li:2022zjc, Hashimoto:2022eij, Ahn:2024gjf, Luo:2024iwf, Chen:2024ckb, Gu:2024lrz, Chen:2024mmd, Cai:2024eqa, Ahn:2024jkk, Mansouri:2024uwc, Chen:2024epd, Hashimoto:2024yev, Ahn:2025tjp, Ran:2025vat}. Collectively, these developments provide profound insights into
how boundary information encodes the bulk spacetime in holographic scenarios.

Previously, our reconstruction scheme~\cite{Lu:2025pal, Lu:2025jgk} was confined to maximally symmetric planar black holes, a simplified setting involving only two independent metric functions.  In this paper, we investigate the extent to which this method relies on spacetime symmetries. We begin by considering black holes with maximally symmetric horizons (namely, planar, spherical, and hyperbolic topologies) and demonstrate that the reconstruction scheme remains valid. More significantly, we achieve the application of this framework to rotating black holes. This generalization is non-trivial because rotation introduces off-diagonal metric components, thereby increasing the number of independent metric functions to be reconstructed.  For three-dimensional rotating black holes, there are three independent metric functions to be reconstructed; we demonstrate that the scheme remains valid in this context. To illustrate the procedure, we explicitly compute the near-horizon metric derivatives for the rotating BTZ black hole~\cite{Banados:1992wn, Banados:1992gq}.  

However, for four-dimensional rotating black holes, the wave equation of a probe scalar field is generically coupled, requiring the solution of partial differential equations that depend on both radial and angular coordinates. In this work, we restrict our study to spacetimes admitting a separable coordinate system~\cite{Carter1968}. This property ensures that the wave equation decouples into a pair of ordinary differential equations (one radial and one angular), while the spacetime geometry is characterized by four unknown metric functions: two radial-dependent and two angular-dependent components. We demonstrate that the radial-dependent metric functions can be reconstructed from pole-skipping data extracted via near-horizon analysis, analogous to the static case. In contrast, the angular-dependent components remain inaccessible from this specific dataset.

To complete the reconstruction of the angular components, we propose a mathematical analogue of pole skipping, termed ``angular pole-skipping,''\footnote{A detailed discussion of this concept is deferred to section \ref{sec:angular_ps}.}, which is defined via a near-axis expansion of the angular differential equation in analogy to the radial case. We show that these angular-dependent components of the metric can be successfully recovered from the angular pole-skipping data. 
Complementing the radial results, these findings provide a comprehensive map for recovering the full spacetime geometry throughout the analytic domain of the metric functions, a framework that we ultimately apply to the Kerr-Newman-AdS black hole.

Moreover, by utilizing the reconstructed metric derivatives at the horizon, we reinterpret the vacuum Einstein equations as a series of algebraic equations in the pole-skipping data. We further demonstrate that the Null Energy Condition (NEC), a fundamental requirement in classical gravity, can impose algebraic inequalities on this data. Finally, we derive general polynomial constraints that arise from the overdetermined structure of our reconstruction scheme.

The paper is organized as follows. In section~\ref{Remaxbh}, we review the reconstruction scheme for static topological black holes using Schwarzschild-like coordinates. In section~\ref{3Drotating}, we extend this method to three-dimensional rotating spacetimes and verify it with the BTZ solution. Section~\ref{Rc4D} deals with the more complex four-dimensional rotating case, utilizing the separability of the wave equation. In section~\ref{EENEC}, we reformulate the vacuum Einstein equations as algebraic equations for the pole-skipping parameters, and extract an algebraic inequality directly from the NEC. Section~\ref{sec:universality} discusses the general structure of the constraints arising from the overdetermined structure of the problem. We conclude with a discussion in section~\ref{sec:discussion}.

\section{\texorpdfstring{Static topological black holes}{Static topological black holes}}\label{Remaxbh}
In contrast to previous works~\cite{Lu:2025pal, Lu:2025jgk}, which employed Eddington-Finkelstein coordinates to ensure manifest regularity across the horizon, we adopt Schwarzschild-like coordinates here. This choice is convenient for handling the complex off-diagonal terms induced by rotation in later sections.

\subsection{Setup and reconstruction}\label{sec:recon_scheme}

We consider a general static \((d+2)\)-dimensional black hole spacetime with maximally symmetric spatial sections, described by the ansatz
\begin{equation}\label{anst1}
d s^2=\frac{1}{z^2}\left(-f_1(z)\,d t^2 + \frac{d z^2}{f_2(z)} + d \Sigma_\kappa^2\right)\,.
\end{equation}
Here, \(\kappa=0, \pm 1\) labels the planar, spherical, and hyperbolic horizon geometries, respectively.
The metric ansatz admits a horizon at \(z=z_h\) defined by \(f_1(z_h) = f_2(z_h) = 0\), while the boundary, corresponding to the asymptotic infinity of the spacetime, is located at \(z=0\). The Hawking temperature of the black hole is given by
\begin{equation}\label{HT1}
T = \frac{1}{4\pi}\sqrt{f_1'(z_h)f_2'(z_h)}\,.
\end{equation}

We introduce a free scalar field $\Phi$ with mass $m$ to probe this background, neglecting its backreaction. The field dynamics are governed by the Klein–Gordon (KG) equation
\begin{equation}\label{KG}
\big(\Box - m^2\big)\Phi = 0 \,.
\end{equation}
Given the spatial symmetry of the metric \eqref{anst1}, we perform a separation of variables for the scalar field
\begin{equation}
\Phi(t,z,X)
=e^{-i\omega t}\,\psi_{\omega,\mu}(z)\,Y_{\mu}(X)\,,
\end{equation}
where $Y_{\mu}(X)$ represents the harmonics on the spatial section $\Sigma_\kappa$, satisfying the eigenvalue equation for the spatial Laplacian $\Delta_{\Sigma}Y_{\mu}=-\mu Y_{\mu}$. For a physical wave propagating in the maximally symmetric black hole spacetime, the eigenvalue \(\mu\) is given by
\begin{align}
\mu =
\begin{cases}
k^2, & \mathbb R^d, \\
\ell(\ell+d-1), &S^d, \\
(\frac{d-1}{2})^2+\nu^2, & H^d,
\end{cases}
\end{align}
with \(\ell=0,1,2,\dots\), and \(\nu\ge0\). In this work, however, we treat the parameter $\mu$ as a generic complex variable. This extension is essential because pole-skipping points typically correspond to complex momenta, which lie outside the standard physical spectrum \cite{Grozdanov:2023txs}.

Upon separating variables in the KG equation \eqref{KG}, we observe that all three transverse geometries yield an identical radial equation
\begin{equation}\label{EoM1}
\psi''+\left(-\frac{d}{2}+\frac{f_1'}{2f_1}+\frac{f_2'}{2f_2}\right)\psi'+\left(\frac{\omega^2}{f_1f_2}-\frac{m^2+\mu z^2}{z^2f_2} \right)\psi = 0\,.
\end{equation}
Since the differential equation \eqref{EoM1} exhibits a regular singularity at the horizon \(z=z_h\), the solution \(\psi\) can be expanded near the horizon as a Frobenius series:
\begin{equation}\label{field1}
\psi(z) = \big(z - z_h\big)^{\rho} \sum_{n=0}^{\infty}\frac{\phi_n }{n!} \big(z - z_h\big)^n,
\end{equation}
where \(\rho\) is the characteristic exponent. For generic  \(\omega\), the near-horizon analysis yields two linearly independent solutions characterized by the exponents \(\rho_\pm=\pm i\omega/(4\pi T)\), which correspond to outgoing and ingoing modes, respectively.

To proceed, we assume that the unknown metric functions admit a Taylor expansion near the horizon
\begin{equation}\label{fa1}
 f_a(z) = \sum_{n=1}^{\infty} \frac{f_a^{(n)}(z_h)}{n!}\,(z - z_h)^n,
 \quad a = 1,2,
\end{equation}
and introduce the shorthand notation
\(
f_{an} \equiv f_a^{(n)}(z_h).
\)

The validity of the reconstruction via the Taylor expansion \eqref{fa1} is mathematically governed by the radius of convergence. For the exact solutions studied in this section (such as Schwarzschild-AdS and RN-AdS), the metric functions $f_a(z)$ are polynomials, ensuring that the series converges in both the exterior and interior regions. However, for more general static backgrounds compatible with the ansatz \eqref{anst1}, the radius of convergence is limited by the nearest singularity in the complex plane.

Proceeding with this series representation to solve the KG equation, we substitute the metric expansion \eqref{fa1}, the scalar field \eqref{field1} (with the ingoing exponent) and the Hawking temperature \eqref{HT1} into \eqref{EoM1}. Expanding the result near the horizon yields the series equation
\begin{equation}\label{S1}
0 \;=\;
\sum_{n=1}^{\infty}
\mathcal{S}_{n}\bigl(f_{aj},\,\phi_j,\,\omega,\,\mu\bigr)\,(z - z_h)^{n-1}.
\end{equation}
For generic $\omega$, requiring that the series equation \eqref{S1} holds for all \(z\) within the convergence radius forces each coefficient \(\mathcal{S}_{n}\) to vanish independently. This yields a tower of linear constraints, which recursively determine the horizon expansion coefficients \(\phi_i \).

Motivated by the prospect of holographic bulk reconstruction, we formulate this as an inverse problem. We assume access to the complete pole-skipping dataset
\begin{equation}\label{ps1}
\mathcal{D} \;=\;
\bigl\{\,(\omega_n,\;\bm\mu_n)\;\bigm|\;n\in\mathbb{Z}_{>0}\bigr\},\quad
\omega_n = -\,i\,2\pi\,T_b\,n,
\quad
\bm\mu_n =\{\mu_{n1},\,\dots,\,\mu_{nn}\},
\end{equation}
where \(T_b\) denotes the boundary temperature,\footnote{We work in a setup where the boundary temperature is identified with the Hawking temperature: \(T_b=T\).} \(\bm{\mu}_n\) represents the set of generalized momentum squared values \(\mu_{n i}\) characterizing the pole-skipping points at order \(n\) and \(\omega_n\) corresponds to the negative imaginary Matsubara frequency. Our objective is to reconstruct the bulk metric functions \(f_1(z)\) and \(f_2(z)\) solely from this data \(\mathcal{D}\).

In the context of the dual field theory, these data correspond to specific points in the complex $(\omega, \mu)$ plane where the retarded Green's function $G_R(\omega, \mu)$ of the scalar operator is ill-defined. These ambiguities are holographic signatures that allow us to probe the bulk geometry.

At these specific frequencies, the prefactor of the $n$th-order expansion coefficient $\phi_n$ vanishes, rendering it a free parameter. The first \(n\) equations in~\eqref{S1} (i.e., \(\mathcal{S}_1 = \dots = \mathcal{S}_n = 0\)) decouple from the higher-order terms. They form a closed \(n\times n\) homogeneous linear system for the vector \(\tilde{\phi}=(\phi_0,\dots,\phi_{n-1})^T\)~\cite{Blake:2019otz}
\begin{equation}
 \mathcal{M}^{(n)}(\{f_{aj}\}_{j\le n},\,\omega_n,\mu_{ni})\cdot \tilde{\phi} =0\,.
\end{equation}
Since the term involving $\phi_n$ vanishes, the system constrains only the lower-order coefficients $\tilde{\phi}$. For a non-trivial solution to exist, the determinant of this coefficient matrix must vanish
\begin{equation}\label{dtM1}
\det \mathcal{M}^{(n)}\bigl(\{f_{aj}\}_{j\le n},\,\omega_n,\,\mu_{ni}\bigr)=0\,,
\end{equation}
This algebraic constraint is satisfied by a set of $n$ values $\bm\mu_n = \{\mu_{n1}, \dots, \mu_{nn}\}$. Specifically, the determinant condition \eqref{dtM1} manifests as a polynomial equation of degree $n$ in $\mu$
\begin{equation}\label{pon1}
V_{n,n} \mu^n +V_{n,n-1} \mu^{n-1}+ \cdots + V_{n,0}=0,
\end{equation}
where the coefficients \(V_{n,i}=V_{n, i}\big(\{f_{aj}\}_{j\le n},\omega_n\big)\) ($i=0,1,2,...,n$) depend on the metric derivatives and the pole-skipping frequency. The \(n\) roots of this polynomial are precisely the elements of the set \(\bm{\mu}_n\).

By Vieta's formulas, the coefficients of the polynomial  \(V_{n,i}\) can be expressed in terms of the elementary symmetric polynomials in its roots \(\bm\mu_n\). We define these as\footnote{In the case \(n=1\) and \(k=0\), we set \(E_{1}(\mu^0)=1\).}
\begin{equation}\label{ek1}
E_{n}(\mu^k) = \sum_{1 \leq i_1 < \cdots < i_k \leq n} \mu_{ni_1} \mu_{ni_2} \cdots \mu_{ni_k}, \quad k = 1,\dots,n.
\end{equation}
where the superscript $k$ denotes the degree of the symmetric polynomial.
Notably, the \(n\)th-order derivatives \( f_{an} \) appear linearly and non-trivially only in the two lowest-order terms, \( V_{n,0} \) and \( V_{n,1} \), and are absent from all higher-order coefficients \(V_{n,i} \) ($i\geq 2$) \cite{Lu:2025pal}.

Applying Vieta’s formulas, we obtain the following two linear equations for \( f_{1n} \) and \( f_{2n} \)
\begin{equation}
\begin{split}
 V_{n,0}\big(f_{an}\big) &= (-1)^n V_{n,n}\, E_{n}\big(\mu^n\big), \\[6pt]
 V_{n,1}\big(f_{an}\big) &= (-1)^{n-1} V_{n,n}\, E_{n}\big(\mu^{n-1}\big). \label{VF1}
\end{split}
\end{equation}
Given the coefficients \( f_{ai} \) with \( i < n \), these two linear equations suffice to determine \( f_{an} \), thereby enabling a recursive construction of the full solution.
\subsection{Explicit expressions of metric derivatives}\label{sec:iter_static}
We now solve the system \eqref{VF1} order by order to determine the metric derivatives up to the third order.

At order \(n=1\), the two equations are linearly dependent, reducing to a single independent constraint\footnote{We set the location of the horizon to $z_h=1$ in this paper. Furthermore, we set the mass $m=0$ in this section for simplicity.}
\begin{equation}
    2d+\frac{f_{11}\,E_1(\mu)}{\omega_1^2}=0.
\end{equation}
Solving it gives the first nontrivial coefficient
\begin{equation}\label{f11}
f_{11}=-\frac{2d\,\omega_1^2}{E_{1}(\mu)},
\end{equation}
where \(E_{1}(\mu)=\mu_{11}\). The second coefficient \(f_{21}\) is then fixed by the Hawking temperature definition \eqref{HT1}
\begin{equation}\label{f21}
 f_{21}=\frac{16\pi^2T^2}{f_{11}} = \frac{-4\omega_1^2}{f_{11}} = \frac{2E_{1}(\mu)}{d}\,,
\end{equation}
where we used \(\omega_1 = -2\pi i T\).

At order \(n=2\), we substitute the result of the first-order derivative \eqref{f11} into system \eqref{VF1}, yielding
\begin{equation}
    \begin{split}
        4d+\frac{E_1(\mu)\,f_{12}}{\omega_1^2}+\frac{d^2\,E_2(\mu^2)}{E_1(\mu)^2}-\frac{d^2E_1(\mu)\,f_{22}}{E_1(\mu)^2}&=0\,,\\[6pt]
        \frac{3f_{12}}{\omega_1^2}+\frac{d^2\,\big(4E_{2}(\mu)-8E_1(\mu)\big)}{E_1(\mu)^2}-\frac{d^2\,f_{22}}{E_1(\mu)^2}&=0\,.
    \end{split}
\end{equation}
Solving these two equations, we obtain the explicit expressions as follows:
\begin{equation}\label{f212}
 \begin{split}
 f_{12}&=\frac{d\,\omega_1^2}{2E_{1}(\mu)^3}\Big[(4+8d)E_1(\mu)^2+d\,E_2(\mu^2)-4d\,E_2(\mu)\,E_1(\mu)\Big]\,,\\[6pt]
f_{22}&=-2E_2(\mu)+\frac{3E_2(\mu^2)}{2E_1(\mu)}+\frac{(4d+6)E_1(\mu)}{d}\,.
\end{split}
\end{equation}

Similarly, at third order ($n=3$), substitution of the lower-order results \eqref{f11} and \eqref{f212} into \eqref{VF1} leads to the linear equations for $f_{a3}$. Solving these equations, we obtain the following explicit expressions
\begin{equation}
    \begin{split}
    f_{13}&=
 -\frac{d\, \omega_1^2}{24 E_1(\mu)^5} \bigg[9 d^2 E_2(\mu^2)^2 - 36 d^2 E_2(\mu) E_2(\mu^2) E_1(\mu) - 2 d^2 E_3(\mu^3) E_1(\mu) \\[4pt]
 &\qquad\qquad+ 
 36 d\, E_2(\mu^2) E_1(\mu)^2 + 18 d^2 E_3(\mu^2) E_1(\mu)^2 - 144 d\, E_2(\mu) E_1(\mu)^3 \\[4pt]
 &\qquad\qquad- 
 72 d^2 E_2(\mu) E_1(\mu)^3 + 96 E_1(\mu)^4 + 288 d\, E_1(\mu)^4 + 
 156 d^2    E_1(\mu)^4
\bigg]\,,
    \end{split}
\end{equation}
and
\begin{equation}
\begin{split}
f_{23} &=  \frac{d}{24 E_1(\mu)^3} \bigg[
 144 E_2(\mu^2) E_1(\mu)^2- 36 E_2(\mu) E_2(\mu^2) E_1(\mu) + 10 E_3(\mu^3) E_1(\mu) \\[4pt]
&\qquad\qquad+ 72 E_2(\mu)^2 E_1(\mu)^2 -9 E_2(\mu^2)^2  - 18 E_3(\mu^2) E_1(\mu)^2 + 
 156 E_1(\mu)^4\\[4pt]
 &\qquad\qquad
 - 216 E_2(\mu) E_1(\mu)^3\bigg]+12 E_1(\mu) - 6 E_2(\mu)+\frac{4 E_1(\mu)}{d} + \frac{9 E_2(\mu^2)}{2 E_1(\mu)} .
\end{split}
\end{equation}

These results demonstrate that the metric derivatives can be systematically reconstructed order by order. Although the algebraic expressions become increasingly cumbersome at higher orders, the procedure remains straightforward. 

Typically, our reconstruction process simplifies because we deliberately isolate the conformal factor $1/z^2$ in the ansatz~\eqref{anst1}, thereby rendering the functions $f_a(z)$ as polynomials.
Specifically, in well-known black hole backgrounds, such as Schwarzschild-AdS and RN-AdS, the higher-order derivative coefficients $f_{an}$ vanish for sufficiently large $n$. Consequently, although the exact reconstruction of $f_a(z)$ formally involves derivatives to all orders, the series effectively truncates at a finite order.

\subsection{Example: the RN-AdS black hole}
\label{sec:RN_example}

To illustrate and verify our reconstruction scheme presented in section~\ref{sec:recon_scheme}, we consider the Reissner-Nordström-AdS$_4$ black hole. This corresponds to the case \(d=2\) in our general formalism. For this solution, the metric functions are identical,
\begin{equation}
\label{RN_f}
f_1(z) = f_2(z) \equiv f(z) = 1 - M z^3 + Q^2 z^4,
\end{equation}
where \(M\) and \(Q\) are the mass and charge, respectively. The event horizon \(z_h\) is defined by \(f(z_h)=0\). By normalizing the radial coordinate to $z_h=1$, we constrain the mass by the charge via
\begin{equation}
\label{eq:RN_MQ_relation}
M = 1 + Q^2.
\end{equation}
The corresponding Hawking temperature is
\begin{equation}
\label{eq:hawking_temp}
T = -\frac{f'(z_h)}{4\pi} = \frac{ 3 - Q^2}{4\pi}.
\end{equation}
For the exact solution \eqref{RN_f}, the theoretical derivatives  are
\begin{equation} \label{eq:RN_theoretical_derivs}
\begin{aligned}
&f^{(1)}(z_h) = Q^2-3, \quad & f^{(2)}(z_h) &= 6 (Q^2-1), \\
&f^{(3)}(z_h) = 18 Q^2-3, \quad & f^{(4)}(z_h) &= 24 Q^2, \\
&f^{(n \ge 5)}(z_h) = 0.
\end{aligned}
\end{equation}
This explicitly demonstrates that the metric derivatives naturally terminate at a finite order in our ansatz. Consequently, the task of reconstructing the full metric reduces to determining the values of these few non-vanishing derivatives.

To extract these derivatives efficiently, instead of determining the individual pole-skipping points $\mu_{ni}$, we directly compute the elementary symmetric polynomials $E_n(\mu^k)$, whose relationship to $\mu_{ni}$ is defined in \eqref{ek1}. By induction, we derive the general formula for $E_n(\mu)$, which reads:
\begin{equation}
    E_n(\mu)=4\pi T n^3+2n(n^2-1).
\end{equation}

For the higher-order terms $E_n(\mu^k)$ with $k>1$, although a simple closed-form expression is not available, we can calculate these terms by substituting the theoretical derivatives \eqref{eq:RN_theoretical_derivs} directly into Vieta's formulas:
\begin{equation}
    E_n(\mu^k)=(-1)^k\frac{V_{n,n-k}}{V_{n,n}},
\end{equation}
where the coefficients $V_{n,i}$, defined in \eqref{pon1}, are constructed from these derivatives. The explicit expressions for these symmetric polynomials at the first few orders are presented in appendix~\ref{data:RN}

By inserting the calculated $E_n(\mu^k)$ into the reconstruction scheme of section~\ref{sec:iter_static}, we find that the reconstructed coefficients are in exact agreement with the theoretical predictions.

\section{Rotating black holes in three dimensions}\label{3Drotating}

\subsection{Setup and reconstruction}

We now generalize the preceding analysis from static spacetimes \eqref{anst1} to rotating black holes. The inclusion of angular momentum breaks the spatial symmetry, thereby increasing the number of independent metric components. We consider an ansatz for three-dimensional rotating black holes that preserves stationarity and axial symmetry
\begin{equation}\label{3rBH}
ds^2=-\frac{f_1(z)}{z^2}\,d t^2+ \frac{d z^2}{z^2f_2(z)} +\frac{1}{z^2}\big(d\varphi-f_3(z)\,d t\big)^2 \,.
\end{equation}
The angular coordinate \(\varphi\) is periodic with \(\varphi \sim \varphi + 2\pi\), and the metric admits an outer horizon at \(z=z_h\), defined by \(f_1(z_h) = f_2(z_h) = 0\). The Hawking temperature is given by \eqref{HT1}. The angular velocity of the horizon is \(\Omega_H = f_3(z_h)\).

Since the metric ansatz \eqref{3rBH} involves three independent functions \(\{f_1, f_2, f_3\}\), we must generalize the reconstruction scheme. We again use a free scalar field \(\Phi\) with mass \(m\), which decomposes as
\begin{equation}
 \Phi(t,z,\varphi) =e^{-i\omega t+i k \varphi}\psi(z)\,.
\end{equation}
Performing a separation of variables in the KG equation yields the radial equation
\begin{equation}\label{EOM2}
\psi''+\bigg(\frac{f_2'}{2f_2} +\frac{f_1'}{2f_1} -\frac{1}{z} \bigg)\psi' +\bigg(\frac{(\omega-k f_3)^2}{f_1\,f_2} -\frac{k^2\,z^2+m^2}{z^2f_2}\bigg)\psi =0\,,
\end{equation}
 The near-horizon expansions for the solution and the metric functions are
\begin{equation}\label{ff}
\begin{split}
\psi(z)& = \big(z - z_h\big)^{\rho} \sum_{n=0}^{\infty}\frac{\phi_{n} }{n!} \big(z - z_h\big)^n\,,\\[6pt]
f_{a}(z)& = \sum_{n=1}^{\infty} \frac{f_{a}^{(n)}(z_h)}{n!}\,(z - z_h)^n,
\quad a = 1,2,\\[6pt]
 f_{3}(z)& = \sum_{n=0}^{\infty} \frac{f_{3}^{(n)}(z_h)}{n!}\,(z - z_h)^n,
\end{split}
\end{equation}
We use the shorthand \(f_{a n} \equiv f_{a}^{(n)}(z_h)\), noting that \(f_{10}=f_{20}=0\) while \(f_{30}=\Omega_H\). The characteristic exponent \(\rho\) for the ingoing mode is
\begin{equation}
\rho \equiv \rho_- = -\frac{i(\omega-k\Omega_H)}{4\pi\, T}\,.
\end{equation}
For simplicity, we introduce the effective frequency \(\mathfrak{w}=\omega-k\Omega_H\). Substituting the expansions \eqref{ff} into the radial equation \eqref{EOM2} yields the series equation
\begin{equation}\label{S2}
0 \;=\;
\sum_{n=1}^{\infty}
\mathcal{S}_{n}\bigl(f_{a i},\,\phi_{ i},\,\mathfrak{w},\,k \bigr)\,(z - z_h)^{n-1}\,.\notag
\end{equation}
As in the static case, pole-skipping occurs at a discrete set of points where the determinant of the \(n\times n\) system for \(\tilde{\phi}=(\phi_0, \dots, \phi_{n-1})^T\) vanishes. This dataset is given by
\begin{equation}\label{ps2}
\mathcal{D} \;=\;
\bigl\{\,(\mathfrak{w}
_{n},\;\bm k_{n})\;\bigm|\;n\in\mathbb{Z}_{>0}\bigr\},\quad
\mathfrak{w}_{n} = -\,i\,2\pi\,T\,n,\quad
\bm k_{n} =\{k_{n1},\dots,k_{n,2n}\}.
\end{equation}
In contrast to the static case \eqref{ps1}, the set $\bm k_n$ contains $2n$ values at each order $n$. The condition \(\det \mathcal{M}^{(n)}=0\) becomes a degree-\(2n\) polynomial in \(k\)
\begin{equation}\label{pon2}
V_{n,2n}k^{2n}+V_{n,2n-1}k^{2n-1}+\cdots+V_{n,1}k + V_{n, 0}=0\,,
\end{equation}
where the \(2n\) roots precisely correspond to the elements of \(\bm{k}_{n}\). We define the elementary symmetric polynomials in these roots as
\begin{align}\label{Ek2}
E_{n}\left(k^l\right) = \sum_{1 \leq i_1 < \cdots < i_l \leq 2n} k_{n i_1} k_{n i_2} \cdots k_{n i_l}, \quad l = 1,\dots,2n.
\end{align}
Crucially, the $n$th order derivatives $f_{an}$ ($a=1,2,3$) enter linearly into only the three lowest-order coefficients $V_{n,0}$, $V_{n,1}$, and $V_{n,2}$; all higher-order coefficients are independent of $f_{an}$ (see appendix~\ref{app_linear_f1f2f3} for a rigorous proof).

 To reconstruct \(f_{an}\), we use the three Vieta's formulas relating the lowest coefficients to the roots
\begin{equation}\label{V2}
\begin{split}
V_{n,0}(f_{an}) &= (-1)^{2n} V_{n,2n}\, E_{n}( k^{2n}), \\[6pt]
V_{n,1}(f_{an}) &= (-1)^{2n-1} V_{n,2n}\, E_{n}(k^{2n-1})\,,\\[6pt]
V_{n,2}(f_{an}) &= (-1)^{2n-2} V_{n,2n}\, E_{n}(k^{2n-2})\,.
 \end{split}
\end{equation}
Given the lower-order coefficients \( f_{ai} \) (\(i < n\)), these three linear equations  allow us to determine the coefficients \( f_{an} \).

\subsection{Explicit expressions of metric derivatives}\label{sec:iter_rotating}
We now determine the metric derivatives order by order by solving Eqs.~\eqref{V2}.

At order \(n=1\), the three equations in \eqref{V2} reduce to two independent constraints\footnote{While the reconstruction presented in section~\ref{sec:iter_static} was restricted to the $m=0$ case for simplicity, the scheme in fact holds for a general mass. In this section, we explicitly generalize it to $m \neq 0$.}
\begin{equation}\label{F1}
\begin{split}
\frac{1}{2}+\frac{\big(m^2-E_1(k^2)\big)f_{11}}{4\mathfrak{w}_1^2}=0\,,\\
\frac{f_{31}}{2\mathfrak{w}_1}+\frac{E_1(k)f_{11}}{4\mathfrak{w}_1^2}=0\,.
 \end{split}
\end{equation}
These two equations determine \(f_{11}\) and \(f_{31}\). The explicit solutions are
\begin{equation}\label{f131}
\begin{split}
 f_{11} &= \frac{2\mathfrak{w}_1^2}{E_{1}(k^2)-m^2}\,,\quad f_{31}=\frac{E_{1}(k)\mathfrak{w}_1}{m^2-E_{1}(k^2)}\,.
\end{split}
\end{equation}
We assume \(E_{1}(k^2)-m^2\neq0\) to ensure regularity. The remaining derivative, \(f_{21}\), is computed using the Hawking temperature formula \eqref{HT1}
\begin{equation}
 f_{21}=\frac{16\pi^2T^2}{f_{11}} = \frac{-4\mathfrak{w}_1^2}{f_{11}} = 2\left(m^2-E_{1}(k^2)\right)\,.
\end{equation}

At order $n=2$, substituting \eqref{f131} into \eqref{V2} yields the following three linear equations for $f_{a2}$
\begin{equation}
\begin{split}
0= &\Big(-4 E_1(k^2)^3 + 9 E_1(k^2)^2 \,m^2 - 6 E_1(k^2)\, m^4 + m^6\Big) 
   f_{12} +\mathfrak{w}_1^2 \Big(4 E_1(k^2) - 3 m^2 \Big)f_{22} \\[4pt]
  &\qquad+ \mathfrak{w}_1^2\Big(16 E_1(k^2)^2  + 4 E_2(k^4)  - 24 E_1(k^2)\, m^2 + 
 4 m^4\Big) \,,\\[6pt]
 0=& \Big(-E_1(k) E_1(k^2)^2 + 2 E_1(k) E_1(k^2)\, m^2 - E_1(k)\, m^4\Big)
   f_{12}  + E_1(k)\, \mathfrak{w}_1^2\, f_{22}\\[4pt] 
  &\qquad+ 
 4 \mathfrak{w}_1\Big(E_1(k^2) - m^2\Big)^2  f_{32}+\mathfrak{w}_1^2\Big(-8 E_1(k) E_1(k^2)  + 
 2 E_2(k^3) \Big) \,,\\[6pt]
 0=& \Big(3 E_1(k^2)^2 - 6 E_1(k^2) m^2 + 3 m^4\Big) f_{12} -
  \mathfrak{w}_1^2 f_{22}+\mathfrak{w}_1^2\Big(12 E_1(k)^2  + 8 E_1(k^2)  - 
 4 E_2(k^2) \Big)\,. 
\end{split}
\end{equation}
Solving this system explicitly, we obtain
\begin{equation}\label{f1232}
\begin{split}
f_{12}=&-\frac{\mathfrak{w}_1^2}{2 \big(E_1(k^2) - m^2\big)^3} \bigg[ m^4+ \Big( 3 E_2(k^2)-9 E_1(k)^2 - 12 E_1(k^2) \Big) m^2+ 12 E_1(k^2)^2  \\
&\qquad\qquad\qquad\qquad\qquad\quad+12 E_1(k)^2 E_1(k^2) - 
 4 E_1(k^2) E_2(k^2) + E_2(k^4) \bigg]\,,\\[6pt]
 f_{22}=&-\frac{1}{2\big(E_1(k^2)-m^2\big)}\bigg[3 m^4 + 
  \Big(E_2(k^2)-3 E_1(k)^2 - 20 E_1(k^2) \Big) m^2 +3 E_2(k^4) \\
  &\qquad\qquad\qquad\qquad\qquad+ 12 E_1(k)^2 E_1(k^2) + 20 E_1(k^2)^2 - 4 E_1(k^2) E_2(k^2)
 \bigg]\,,\\[6pt]
 f_{32}=&\frac{\mathfrak{w}_1}{4\big(E_1(k^2)-m^2\big)^3}\bigg[  \Big(3 E_1(k)^3 - 12 E_1(k) E_1(k^2) - E_1(k) E_2(k^2) + 2 E_2(k
 ^3)\Big) m^2\\
 &\qquad\qquad +E_1(k) m^4 +12 E_1(k) E_1(k^2)^2 - 2 E_1(k^2) E_2(k^3) + 
 E_1(k) E_2(k^4)  \bigg]\,.
\end{split}
\end{equation}
Because the algebraic expressions at the third order ($n=3$) are highly unwieldy, we relegate these results to appendix~\ref{f1233}. We note that higher-order derivatives can, in principle, be derived by solving the corresponding linear equations.

\subsection{Example: the rotating BTZ black hole}
\label{sec:BTZ_example}

As a concrete validation of our reconstruction formalism for three-dimensional rotating spacetimes, we apply the scheme to the rotating BTZ black hole~\cite{Banados:1992wn, Banados:1992gq}. This exact solution serves as an ideal testbed, as both the metric derivatives and the pole-skipping data can be computed analytically.

The metric in Schwarzschild coordinates $(t, r, \varphi)$ is given by
\begin{equation}\label{rBTZr}
ds^2 = -N^2(r) dt^2 + \frac{dr^2}{N^2(r)} + r^2 \big(N^{\varphi}(r) dt + d\varphi\big)^2,
\end{equation}
where the lapse function $N(r)$ and the angular shift $N^{\varphi}(r)$ are defined as
\begin{equation}
N^2(r) = -M + r^2 + \frac{J^2}{4r^2} = \frac{(r^2-r^2_{+})(r^2-r^2_-)}{r^2}, \quad
N^{\varphi}(r) = -\frac{J}{2r^2}.
\end{equation}
Here, the mass $M$ and angular momentum $J$ are parameterized by the inner and outer horizon radii, $r_\pm$, via $M=r_{+}^2+r_{-}^2$ and $J=2r_{+}r_{-}$. The Hawking temperature $T$ and the horizon angular velocity $\Omega_H$ are given by
\begin{equation}
T = \frac{r_{+}^2-r_{-}^2}{2\pi r_{+}}, \quad \Omega_{H} = \frac{J}{2r_{+}^2} = \frac{r_-}{r_+}.
\end{equation}

To map this geometry to our reconstruction ansatz \eqref{3rBH}, we perform the coordinate transformation $r=1/z$. This maps the conformal boundary ($r \to \infty$) to $z=0$ and the outer event horizon ($r=r_+$) to $z_h = 1/r_+$. Under this identification, the ansatz functions $f_a(z)$ take the explicit forms
\begin{equation}\label{fBTZtrue}
\begin{split}
f_1(z) &= f_2(z) \equiv f(z) = \left(1-\frac{z^2}{z_+^2}\right)\left(1-\frac{z^2}{z_-^2}\right), \\
f_3(z) &= \frac{J z^2}{2},
\end{split}
\end{equation}
where $z_\pm = 1/r_\pm$.

For the analytical reconstruction test, we select a specific configuration by fixing the horizon radius $z_h \equiv z_+ = 1$ and the temperature $T=1/(4\pi)$. These conditions constrain the inner horizon to $z_- = \sqrt{2}$ and the angular velocity to $\Omega_H = 1/\sqrt{2}$. The theoretical derivatives of the metric functions at the horizon ($z_h=1$), which serve as the benchmark for our reconstruction, are
\begin{equation}\label{dftrue}
\begin{aligned}
&f^{(1)}(z_h) = -1, \quad & f_3^{(1)}(z_h) &= \sqrt{2}, \\
&f^{(2)}(z_h) = 3, \quad & f_3^{(2)}(z_h) &= \sqrt{2}, \\
&f^{(3)}(z_h) = 12, \quad & f_3^{(n \ge 3)}(z_h) &= 0, \\
&f^{(4)}(z_h) = 12, \quad & f^{(n\ge 5)}(z_h) &= 0.
\end{aligned}
\end{equation}

We determine the pole-skipping data for this background through two methods: near-horizon analysis and the holographic Green's function. The pole-skipping frequencies are located at the negative imaginary Matsubara values $\mathfrak{w}_n = -i 2\pi T n$, which simplify to $\mathfrak{w}_n = -i n / 2$ in our setup. At each order $n$, the near-horizon determinant condition yields $2n$ distinct  pole-skipping momenta, given by the formula
\begin{equation}\label{psrBTZ}
k_{nj} = -i \left( \frac{n}{\sqrt{2}} + 2j - n - 1 \pm \sqrt{1 + m^2} \right), \quad j = 1, \dots, n,
\end{equation}
where $m^2 = \Delta(\Delta-2)$ is the bulk scalar mass squared.
\begin{figure}[t]
    \centering
    \includegraphics[width=0.45\linewidth]{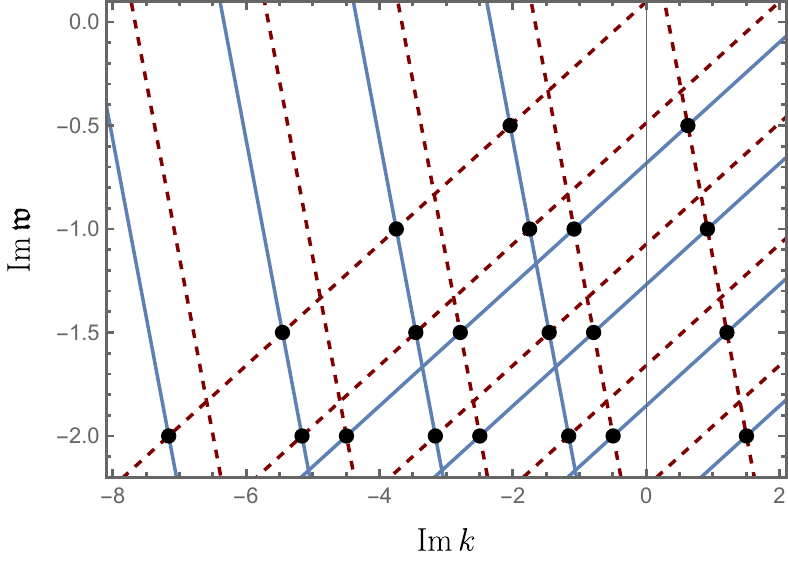}
    \caption{Distribution of pole-skipping points for the rotating BTZ black hole (with $\Delta_+=7/3$). The pole-skipping points appear at the intersections of the curves defined by the poles (solid lines) and zeros (dashed lines) of the retarded Green's function. These intersections precisely match the locations $\big(\mathfrak{m}_n, k_{nj}\big)$, denoted by dots, derived from the near-horizon analysis. Due to the rotation, the distribution of pole-skipping points becomes asymmetric with respect to the imaginary axis.}
    \label{fig:psrBTZ}
\end{figure}
Alternatively, these special points can be identified from the analytic structure of the retarded Green's function in the BTZ background~\cite{Blake:2019otz, Jeong:2023rck}
\begin{equation}
G_{R}(\omega, k) \propto \frac{\Gamma\left(\frac{\Delta_+}{2} - i \frac{\omega+k}{4\pi T_R}\right) \Gamma\left(\frac{\Delta_+}{2} - i \frac{\omega-k}{4\pi T_L}\right)}{\Gamma\left(\frac{\Delta_-}{2} - i \frac{\omega+k}{4\pi T_R}\right) \Gamma\left(\frac{\Delta_-}{2} - i \frac{\omega-k}{4\pi T_L}\right)},
\end{equation}
where $\Delta_{\pm} = 1 \pm \sqrt{1+m^2}$ are the conformal dimensions, and the left/right-moving temperatures are $T_{L/R} = (r_+ \mp r_-)/(2\pi)$. Expressed in terms of the effective frequency near the horizon, $\mathfrak{w} = \omega - \Omega_H k$, the poles and zeros of $G_R$ form families of lines in the complex $(\mathfrak{w}, k)$ plane. Pole-skipping occurs precisely at the intersection points where a pole line crosses a zero line, rendering the Green's function indeterminate ($0/0$). As illustrated in figure~\ref{fig:psrBTZ}, the locations derived from \eqref{psrBTZ} perfectly coincide with these intersection points.

Finally, we input the pole-skipping data derived from \eqref{psrBTZ}—specifically the symmetric polynomials $E_n(k^l)$—into the expressions derived in section~\ref{sec:iter_rotating}. The reconstructed metric coefficients are
\begin{equation} \label{resultBTZ}
\begin{alignedat}{3}
f_{11} &= -1, &\quad f_{21} &= -1, &\quad f_{31} &= \sqrt{2}, \\
f_{12} &= 3, &\quad f_{22} &= 3, &\quad f_{32} &= \sqrt{2}, \\
f_{13} &= 12, &\quad f_{23} &= 12, &\quad f_{33} &= 0, \\
f_{14} &= 12, &\quad f_{24} &= 12, &\quad f_{34} &= 0, \\
f_{1, n \ge 5} &= 0, &\quad f_{2, n \ge 5} &= 0, &\quad f_{3, n \ge 3} &= 0.
\end{alignedat}
\end{equation}
Comparing \eqref{resultBTZ} with the theoretical benchmark \eqref{dftrue}, we observe exact agreement order by order.
\section{Rotating black holes in four dimensions}\label{Rc4D}
We now consider a more intricate and physically relevant scenario of four-dimensional rotating black holes. In this analysis, we explicitly restrict our attention to spacetimes admitting a separable coordinate system. The treatment of non-separable coordinates, which involves coupled partial differential equations, lies beyond the scope of this work.

\subsection{Setup}
In contrast to the previous sections where we employed the $z$-coordinate (with the boundary at $z=0$), here we adopt the standard radial coordinate $r$, where the asymptotic boundary is located at $r \to \infty$. We consider a stationary and axisymmetric Carter-type metric ansatz~\cite{Carter1968} in  $(t,r,x,\varphi)$ coordinates\footnote{Unlike the three-dimensional case where $f_3$ denotes a metric component in the angular sector, here we distinguish the functions such that $f_{1,2}(r)$ are purely radial and $y_{1,2}(x)$ are purely angular.}
\begin{equation}\label{Ct}
 \begin{split}
  ds^2 = &-\frac{f_1(r)}{Z(r,x)}\bigl(dt - y_2(x)\,d\varphi\bigr)^2
  + \frac{y_1(x)}{Z(r,x)}\bigl(a\,dt - f_2(r)\,d\varphi\bigr)^2 \\ 
  &+ \frac{Z(r,x)}{f_1(r)}\,dr^2
  + \frac{Z(r,x)}{y_1(x)}\,dx^2,
 \end{split}
\end{equation}
where \(a\) is the rotation parameter and \(\varphi\) is the periodic azimuthal coordinate. The coordinate $r$ is the radial distance and $x$ represents the cosine of the polar angle (e.g., \(x = \cos\theta \in [-1, 1]\)). This ansatz is well known to possess a hidden symmetry that ensures the separability of the KG equation~\cite{Carter1968}, provided that the function \(Z(r,x)\) is additively separable
\begin{equation}
 Z(r,x) = f_2(r) - a y_2(x).
\end{equation}
The metric~\eqref{Ct}, which encompasses the Kerr family of spacetimes, involves four unknown functions: two radial functions,
$f_{1,2}(r)$, and two angular functions,  $y_{1,2}(x)$.
The spacetime admits an outer event horizon at \(r=r_h\), defined by the root \(f_1(r_h)=0\). The Hawking temperature \(T\) and the horizon's angular velocity \(\Omega_H\) are given by
\begin{equation}\label{Tandve4d}
T=\frac{f_1'(r_h)}{4\pi f_2(r_h)}\,,\quad \Omega_H=\frac{a}{f_2(r_h)}.
\end{equation}
We probe this geometry with a free scalar field \(\Phi\) of mass \(m\). Exploiting the symmetries, we decompose the field as
\begin{equation}\label{4dsol}
\Phi(t, \varphi, x, r) = e^{-i\omega t + i k \varphi} \Theta(x) \psi(r).
\end{equation}
Substituting this ansatz into the KG equation \(\big(\Box - m^2\big)\Phi = 0\) and separating variables yields two independent ordinary differential equations, one radial and one angular, coupled by a separation constant \(\lambda\)
\begin{align}
0=&\psi''+\frac{f_1'}{f_1}\psi'+\bigg(\frac{(ak-\omega f_2)^2}{f_1^2}-\frac{\lambda+m^2f_2}{f_1}\bigg)\psi\,,\label{req}\\
0=&\Theta''+\frac{y_1'}{y_1}\Theta'+\bigg(-\frac{(k-\omega y_2)^2}{y_1^2}+\frac{\lambda+am^2y_2}{y_1}\bigg)\Theta.\label{seq}
\end{align}

The pole-skipping analysis for the radial part \eqref{req} follows the same procedure as in the static case. However, unlike the static geometry, the system now involves three independent parameters: the frequency $\omega$, the azimuthal wavenumber $k$, and the separation constant $\lambda$. To systematically extract the pole-skipping points, we treat $k$ as a fixed parameter. Consequently, evaluating the near-horizon system at the special Matsubara frequencies reduces the vanishing determinant condition to a polynomial equation strictly in terms of $\lambda$. The discrete roots of this polynomial will subsequently serve as the radial pole-skipping data for reconstructing the radial metric functions.
\subsection{Angular pole-skipping}\label{sec:angular_ps}
We now consider the angular equation~\eqref{seq}, defined on the physical domain bounded by the regular singular points $x_0$ satisfying $y_1(x_0)=0$. Geometrically, these points lie on the rotation axis. To ensure the spatial geometry closes smoothly at the poles—specifically, that the azimuthal circle shrinks to a point ($g_{\varphi\varphi}\to 0$)—the metric function $y_2(x)$ is required to vanish at the boundaries
\begin{equation}
y_2(x_0) = 0.
\end{equation}

We proceed in strict analogy with the near-horizon analysis used to derive pole-skipping points. We examine the expansion behavior of the angular equation \eqref{seq} near the rotation axis by performing a series expansion for both the scalar field and the metric functions
\begin{equation}\label{expfm}
\begin{split}
\Theta(x)&=\big(x- x_0\big)^{\rho} \sum_{n=0}^{\infty}\frac{\Theta_{n} }{n!} \big(x - x_0\big)^n\,,\\[6pt]
y_{\alpha}(x)& = \sum_{n=1}^{\infty} \frac{y_{\alpha n}}{n!}\,(x - x_0)^n,
\quad \alpha = 1,2\,.
 \end{split}
\end{equation}
We adopt the shorthand \(y_{\alpha n} \equiv y_\alpha^{(n)}(x_0)\). The expansions for \(y_\alpha\) start at \(n=1\), consistent with the conditions \(y_1(x_0)=y_2(x_0)=0\). The indicial equation for \eqref{seq} yields exponents \(\rho_{\pm}=\pm k/y_{11}\). 

Substituting \eqref{expfm} into the angular equation \eqref{seq} yields a tower of recurrence relations for the coefficients \(\Theta_n\). This system can be compactly expressed in matrix form as
\begin{equation}\label{seri_matrix}
\begin{pmatrix}
M_{11} & y_{11}-2k & 0 & \cdots & 0 \\
M_{21} & M_{22} & 2y_{11}-2k & \cdots & 0 \\
\vdots & \vdots & \vdots & \ddots & \vdots \\
M_{n1} & M_{n2} & \cdots & M_{nn} & ny_{11}-2k
\end{pmatrix}
\begin{pmatrix}
\Theta_0 \\
\Theta_1 \\
\vdots \\
\Theta_n
\end{pmatrix}
= 0 \,.
\end{equation}
At the discrete values of the azimuthal quantum number, 
\begin{equation}\label{kn}
k=k_n=\frac{ny_{11}}{2}\,, \quad n \in \mathbb{Z}_{>0},
\end{equation}
the coefficient of \(\Theta_n\) in the \(n\)th equation vanishes. This has two immediate consequences, strictly analogous to the radial pole-skipping analysis:
\begin{enumerate}[label=(\roman*)]
  \item The coefficient \(\Theta_n\) becomes a free parameter, undetermined by the lower-order coefficients.
  \item The first \(n\) equations decouple from the higher-order terms, forming a closed \(n \times n\) linear system for the vector \(\tilde{\Theta}=(\Theta_0,\dots,\Theta_{n-1})^T\), which we write as \(\mathcal{M}^{(n)}_{\text{ang}}(k_n,\lambda)\cdot \tilde{\Theta}=0\).
\end{enumerate}
The existence of a non-trivial solution \(\tilde{\Theta}\) requires the determinant of the system's matrix to vanish
\begin{equation}
\label{eq:quantization_condition}
\det\mathcal{M}^{(n)}_{\text{ang}}(k_n, \lambda) = 0.
\end{equation}
For a fixed $k_n$, this determinant condition forms a polynomial equation of degree $n$ in the separation constant $\lambda$. We denote the set of its $n$ roots as $\bm{\lambda}^{(a)}_{n} = \{\lambda^{(a)}_{ni}\}_{i=1}^n$, where the superscript $(a)$ denotes the angular sector. Although the physical origin differs from the radial case, the algebraic structure of the data $(k_n, \bm{\lambda}^{(a)}_{n})$ is analogous to that of the conventional radial pole-skipping points $(\omega_n, \bm\mu_n)$. In light of this similar algebraic structure, we refer to the pairs $(k_n, \bm{\lambda}^{(a)}_{n})$ as ``angular pole-skipping.''

\subsection{Reconstruction}\label{sec:recon_4d}
The reconstruction of the four metric functions $\{f_1, f_2, y_1, y_2\}$ decouples into two independent iterative problems: one for the radial functions at the horizon $r_h$, and one for the angular functions at the axis $x_0$.
\paragraph{Radial sector}
The reconstruction of $f_1(r)$ and $f_2(r)$ utilizes the radial pole-skipping data $\mathcal{D}^{(r)}$.
The dataset is defined as
\begin{equation}\label{ps4d_rev}
\mathcal{D}^{(r)} \;=\;
\bigl\{\,\big(\mathfrak{w}_n, \;\bm\lambda^{(r)}_n\big)\;\bigm|\;n\in\mathbb{Z}_{>0}\bigr\},\quad
\bm\lambda^{(r)}_n =\{\lambda^{(r)}_{n1},\,\dots,\,\lambda^{(r)}_{nn}\},
\end{equation}
where $\mathfrak{w}_n=\omega_n-k \Omega_H$ and $\bm\lambda^{(r)}_n$ are the $n$ roots of the radial degeneracy condition
\begin{equation}\label{pon4d_rev}
V^{(r)}_{n,n} \lambda^n +V^{(r)}_{n,n-1} \lambda^{n-1}+ \cdots + V^{(r)}_{n,0}=0\,.
\end{equation}
The coefficients $V^{(r)}_{n,i}$ depend on the lower-order metric derivatives and linearly on the $n$th-order derivatives $f_{1n}$ and $f_{2n}$.
Following the definition in \eqref{ek1}, we denote the elementary symmetric polynomials of the radial eigenvalues as $E^{(r)}_{n}(\lambda^k)$.
Applying Vieta's formulas to the two lowest-order terms in \eqref{pon4d_rev}, we obtain the linear constraints
\begin{equation}\label{VF4d_rev}
\begin{split}
 V^{(r)}_{n,0}\big(f_{an}\big) &= (-1)^n V^{(r)}_{n,n}\, E^{(r)}_{n}\big(\lambda^n\big), \\[6pt]
 V^{(r)}_{n,1}\big(f_{an}\big) &= (-1)^{n-1} V^{(r)}_{n,n}\, E^{(r)}_{n}\big(\lambda^{n-1}\big).
\end{split}
\end{equation}
Together with the Hawking temperature definition \eqref{Tandve4d}, this system can determine $f_{1n}$ and $f_{2n}$.
\paragraph{Angular sector}
The reconstruction of the angular functions $y_1(x)$ and $y_2(x)$ mirrors the radial procedure. We utilize the angular pole-skipping data $\mathcal{D}^{(a)} = \{\big( k_n,\;\bm\lambda^{(a)}_n\big)\}_{n>0}$, 
where $\bm\lambda^{(a)}_n$ denotes the set of roots of the angular determinant equation given in \eqref{eq:quantization_condition}.

Defining the elementary symmetric polynomials $E^{(a)}_{n}(\lambda^k)$ from these roots, the application of Vieta's formulas yields the corresponding linear constraints for the angular derivatives $y_{1n}$ and $y_{2n}$
\begin{equation}\label{VF3_rev}
\begin{split}
 V^{(a)}_{n,0}(y_{\alpha n}) &= (-1)^n V^{(a)}_{n,n}\, E^{(a)}_{n}(\lambda^n), \\[6pt]
 V^{(a)}_{n,1}(y_{\alpha n}) &= (-1)^{n-1} V^{(a)}_{n,n}\, E^{(a)}_{n}(\lambda^{n-1}).
 \end{split}
\end{equation}
This system, supplemented by the regularity condition $y_2(x_0)=0$ (implicit in the expansion starting at $n=1$), allows us to extract the angular metric functions from data $\mathcal{D}^{(a)} $.

\paragraph{Separation constant $\lambda$}
Before proceeding to the explicit computation of the metric derivatives, some remarks on the separation constant $\lambda$ are in order. 

Conventionally, in physical analysis on a rotating background (e.g., Kerr-AdS black hole), the requirement of regularity at the poles ($y_1(x_0)=0$) in~\eqref{seq} discretizes the separation constant $\lambda$, and the corresponding regular eigenfunctions are known as spheroidal harmonics~\cite{ Berti:2005gp}. For a given $\omega$ and integer $k$, the separation constant $\lambda$ at which regular solutions exist can be indexed by an integer $l$, where regularity constrains $ -l\leq k\leq l$ and $l\geq 0$. In general, $\lambda(\omega,k,l)$ must be computed numerically \cite{Cardoso:2013pza}. In the present work, however, we relax the regular boundary conditions and treat $\lambda$ as a generic complex variable. As pointed out in~\cite{Blake:2021hjj}, this treatment is necessary because the pole-skipping points generally correspond to complex values of the separation constant.

Furthermore, to obtain the solution \eqref{4dsol} or quasinormal modes, \eqref{req} and \eqref{seq} should share a common separation constant $\lambda(\omega,k)$, which couples the angular and radial sectors. In this work, however, our goal is to find sufficient data to reconstruct both the radial metric functions $ f(r)$ and angular metric functions $ y(x)$. In general, the pole-skipping datasets $\mathcal{D}^{(r)}$ and $\mathcal{D}^{(a)}$ derived from the radial and angular sectors are mutually disjoint. The special points defining the radial data $\mathcal{D}^{(r)}$ typically manifest as ordinary points in the angular equation, and vice versa. This disjointness arises because, despite their structural similarities, equations \eqref{req} and \eqref{seq} are governed by independent metric functions ($f(r)$ and $y(x)$). Consequently, their respective pole-skipping spectra for the separation constant do not naturally coincide, yielding
\begin{equation}
    \bm\lambda_n^{(r)}\neq \bm\lambda_n^{(a)}.
\end{equation}
To obtain a complete reconstruction, we can use both $\mathcal{D}^{(r)}$ and $\mathcal{D}^{(a)}$ as separate sources of data. This requires us to treat \eqref{req} and \eqref{seq} as two independent equations in our inverse problem, allowing us to determine all four metric functions.

Crucially, at each order $n$, the two spectra $\bm\lambda_n^{(r)}$ and $\bm\lambda_n^{(a)}$ are parameterized by distinct free variables. Specifically, the $n$ radial spectral branches are functions of the azimuthal wavenumber $k$, taking the form $\lambda^{(r)}_{ni}(\mathfrak{w}_n, k)$. In contrast, the angular branches are governed by the effective frequency $\mathfrak{w}$, expressed as $\lambda^{(a)}_{ni}(\mathfrak{w}, k_n)$, where $i = 1, \dots, n$ in both sectors. There is no a priori reason to restrict $k$ in the radial sector or $\mathfrak{w}$ in the angular sector to specific values. Nevertheless, we can utilize these spectral branches to reconstruct the metric.

In practice, to streamline the reconstruction of the metric derivatives, we implement a double-locking prescription
at each order $n$: we evaluate both the radial and angular sectors at the same specific point $(\mathfrak{w}_n, k_n)$ in the parameter space. While this forces the separation constant $\lambda$ to take discrete values, the resulting radial spectrum $\{\lambda_{ni}^{(r)}\}$ and angular spectrum $\{\lambda_{ni}^{(a)}\}$ (where $i=1,\dots,n$) remain distinct, effectively encoding the independent information of the radial and angular metric functions.

It should be stressed that while radial pole-skipping has a well-established holographic dictionary rooted in boundary Green's functions, the physical origin of the angular pole-skipping phenomenon remains an open question. At present, the angular data $\mathcal{D}^{(a)}$ is formulated purely as a bulk-side mathematical counterpart via near-axis expansion. Therefore, it is not yet clear whether the recovery of the angular metric functions $y_{1,2}(x)$ can be strictly classified as a standard boundary-to-bulk holographic reconstruction. Nevertheless, from an algorithmic perspective, this bulk-side formalism is highly non-trivial: it completes the geometric inverse problem, demonstrating that the full rotating spacetime can be analytically determined once the corresponding spectral data $(\lambda_n^{(r)}, \lambda_n^{(a)})$ are provided. We will revisit this issue in section~\ref{sec:discussion}.

\subsection{Explicit expressions of metric derivatives}\label{Itrmdkerr}

We now solve the systems \eqref{VF4d_rev} and \eqref{VF3_rev} for the first few orders in the massless case ($m=0$).

At $n=1$, the equations reduce to independent constraints. Solving them yields the first-order  derivatives
\begin{equation}\label{fy1}
   \begin{split}
     f_{11} &= \frac{2i a\, \mathfrak{w}_1}{\Omega_{H}}, \\
     f_{21} &= \frac{i E^{(r)}_{1}(\lambda)}{\mathfrak{w}_1+k_1\Omega_H},
   \end{split}
\qquad
 \begin{split}
     y_{11} &= 2k_1, \\
     y_{21} &= -\frac{E^{(a)}_{1}(\lambda)}{\mathfrak{w}_1+k_1\Omega_H},
\end{split}
\end{equation}
where $E^{(r)}_{1}(\lambda)=\lambda^{(r)}_{11}$ and $E^{(a)}_{1}(\lambda)=\lambda^{(a)}_{11}$. Note that $f_{11}$ is related to $f_{21}$ by the temperature, and $y_{11}$ is fixed by the regularity condition $k_n = n y_{11}/2$.

At order $n=2$, solving the linear systems explicitly gives
\begin{equation}\label{fy2}
\begin{split}
 f_{12} &= E^{(r)}_{2}(\lambda)-8E_1^{(r)}(\lambda),\quad y_{12} =8E_1^{(a)}(\lambda) -E^{(a)}_{2}(\lambda),\\[6pt]
 f_{22} &= \frac{\Omega_H}{4a\,\mathfrak{w}_1(\mathfrak{w}_1+k_1\Omega_H)}\bigg(E_2^{(r)}(\lambda^2)-2E_2^{(r)}(\lambda)E_1^{(r)}(\lambda)+4E_1^{(r)}(\lambda)^2\bigg),\\[6pt]
 y_{22} &= \frac{1}{4k_1(\mathfrak{w}_1+k_1\Omega_H)}\bigg(E_2^{(a)}(\lambda^2)-2E_2^{(a)}(\lambda)E_1^{(a)}(\lambda)+4E_1^{(a)}(\lambda)^2\bigg).
\end{split}
\end{equation}
Here we have used the notation $E_2(\lambda) = \lambda_{21}+\lambda_{22}$ and $E_2(\lambda^2) = \lambda_{21}\lambda_{22}$, consistent with \eqref{ek1}.

Similarly, at order $n=3$, inserting the known solutions for $n=1,2$ into the recurrence relations generates linear constraints for the third-order derivatives. The explicit solutions for $f_{\alpha 3}$ and $y_{\alpha 3}$ are found to be:
\allowdisplaybreaks[1]
\begin{equation}
    \begin{aligned}
        f_{13} &= \frac{i\Omega_H}{2a\mathfrak{w}_1}\bigg(3 E_2^{(r)}(\lambda)^2 + 9 E_2^{(r)}(\lambda^2) - E_3^{(r)}(\lambda^2) - 12 E^{(r)}_2(\lambda) E^{(r)}_1(\lambda) + 3 E^{(r)}_1(\lambda)^2\bigg)\,, \\[10pt]
        y_{13} &= -\frac{1}{2k_1}\bigg(3 E_2^{(a)}(\lambda)^2 + 9 E_2^{(a)}(\lambda^2) - E_3^{(a)}(\lambda^2) - 12 E^{(a)}_2(\lambda) E^{(a)}_1(\lambda) + 3 E^{(a)}_1(\lambda)^2\bigg)\,, \\[15pt]
        f_{23} &= \frac{i\Omega_H^2} {24a^2\mathfrak{w}_1^2(\mathfrak{w}_1+k_1\Omega_H)}\bigg[ E^{(r)}_3(\lambda^3)-9 E^{(r)}_2(\lambda) E^{(r)}_2(\lambda^2)  + 18 E^{(r)}_2(\lambda)^2 E^{(r)}_1(\lambda) \\
               &\qquad + 36 E^{(r)}_2(\lambda^2) E^{(r)}_1(\lambda) - 3 E^{(r)}_3(\lambda^2) E^{(r)}_1(\lambda) - 72 E^{(r)}_2(\lambda) E^{(r)}_1(\lambda)^2 + 72 E^{(r)}_1(\lambda)^3\bigg], \\[10pt]
        y_{23} &= -\frac{1} {24k_1^2(\mathfrak{w}_1+k_1\Omega_H)}\bigg[E^{(a)}_3(\lambda^3)-9 E^{(a)}_2(\lambda) E^{(a)}_2(\lambda^2)  + 18 E^{(a)}_2(\lambda)^2 E^{(a)}_1(\lambda) \\
               &\qquad + 36 E^{(a)}_2(\lambda^2) E^{(a)}_1(\lambda) - 3 E^{(a)}_3(\lambda^2) E^{(a)}_1(\lambda) - 72 E^{(a)}_2(\lambda) E^{(a)}_1(\lambda)^2 + 72 E^{(a)}_1(\lambda)^3 \bigg].
    \end{aligned}
\end{equation}

\subsection{Example: the Kerr-Newman-AdS black hole}\label{sec:KN_example}
To provide a non-trivial verification of our four-dimensional reconstruction formalism in section~\ref{Itrmdkerr}, we now apply it to the Kerr-Newman-AdS$_4$ (KN-AdS) black hole. The metric in Boyer-Lindquist-type coordinates \((t,r,x,\phi)\), with \(x = \cos\theta\), is given by
\begin{equation}
\begin{split}
   ds^2=&-\frac{\Delta_r(r)}{\Sigma(r,x)}\bigg(dt-\frac{a\big(1-x^2\big)}{\Xi}d\phi \bigg)^2+\frac{\big(1-x^2\big)\Delta_x(x)}{\Sigma(r,x)}\bigg(adt-\frac{r^2+a^2}{\Xi}d\phi \bigg)^2\\[5pt]
   &+\frac{\Sigma(r,x)}{\Delta_r(r)}dr^2+\frac{\Sigma(r,x)}{\big(1-x^2\big)\Delta_{x}(x)}dx^2,
\end{split} 
\end{equation}
where 
\begin{align}
   &\Sigma(r,x)=r^2+a^2x^2,\quad \Xi=1-\frac{a^2}{l^2},\\[6pt]
   &\Delta_r(r)=(r^2+a^2)\left(1+\frac{r^2}{l^2}\right)-2Mr+Q^2,\quad \Delta_x(x)=1-\frac{a^2x^2}{l^2}.
\end{align}
Here \(M\), \(Q\), and \(a\) are the mass, charge, and rotation parameter, respectively. The metric is a solution to the Einstein-Maxwell equations with a cosmological constant \(\Lambda=-3/l^2\). We set the AdS radius \(l=1\) without loss of generality, which requires \(a<1\).

To match our Carter-type ansatz \eqref{Ct}, we rescale the azimuthal coordinate to \(\varphi=\phi/\Xi\). We then identify the metric functions as
\begin{equation}\label{mtKerr}
\begin{aligned}
   Z(r,x)&=\Sigma(r,x), \quad f_1(r)=\Delta_r(r),\quad f_2(r)=r^2+a^2,\\[6pt] y_1(x)&=\big(1-x^2\big)\Delta_x(x),\quad y_2(x)=a\big(1-x^2\big).
\end{aligned}
\end{equation}
We normalize the radial coordinate by setting the outer horizon radius to \(r_h=1\). This fixes the mass parameter via the condition \(\Delta_r(r_h)=0\), yielding
\begin{equation}
\label{eq:KN_MQ_relation}
M =  1 +a^2+ \frac{Q^2}{2}.
\end{equation}
The corresponding Hawking temperature and horizon angular velocity are
\begin{equation}
\label{eq:KN-hawking_temp}
T = \frac{\Delta_r'(r_h)}{4\pi f_2(r_h)} = \frac{4 - Q^2}{4\pi(1+a^2)},\quad \Omega_H=\frac{a}{f_2(r_h)} = \frac{a}{1+a^2}.
\end{equation}

We now proceed to compute the elementary symmetric polynomials \(E_n(\lambda^k)\) for this rotating background. The general formulas for the first-order coefficients in the radial and angular sectors, \(E^{(r)}_n(\lambda)\) and \(E^{(a)}_n(\lambda)\), are derived as
\begin{equation}
\begin{split}
     E^{(r)}_n(\lambda) &= \frac{1}{3}n(n^2-1)(a^2+7) - 4n^3\pi T - i\frac{2n^3a(a^2-1)}{1+a^2}, \\[6pt]
     E^{(a)}_n(\lambda) &= \frac{n}{3(1+a^2)}\Big[ n^2(a^2-1)^2 - 8n^2a^2 + (1+a^2)(5a^2-1) \Big] - 4i n^3 a \pi T.
\end{split}
\end{equation}
For the higher-order symmetric polynomials \(E^{(r)}_n(\lambda^k)\) and \(E^{(a)}_n(\lambda^k)\) with \(k>1\), closed-form expressions are cumbersome. However, they can be determined directly via Vieta's formulas:
\begin{equation}
    E^{(r)}_n(\lambda^k) = (-1)^k \frac{V^{(r)}_{n,n-k}}{V^{(r)}_{n,n}}, \qquad
    E^{(a)}_n(\lambda^k) = (-1)^k \frac{V^{(a)}_{n,n-k}}{V^{(a)}_{n,n}},
\end{equation}
where the coefficients \(V^{(r)}_{n,i}\) and \(V^{(a)}_{n,i}\), defined in \eqref{pon4d_rev} and \eqref{eq:quantization_condition}, are constructed from the derivatives of the metric functions in \eqref{mtKerr}. The
explicit expressions for these symmetric polynomials at the first few orders are presented in appendix~\ref{data:KN}

Finally, inserting the calculated \(E_n^{(r)}(\lambda^k)\) and \(E_n^{(a)}(\lambda^k)\) into the reconstruction scheme of section~\ref{Itrmdkerr} (with \(m=0\)), we find that the reconstructed bulk geometry coefficients are in exact agreement with the theoretical predictions from the metric expansion.

\section{Einstein equations and Null Energy Condition}\label{EENEC}
Having established the metric reconstruction scheme, we now examine how the bulk dynamics imprint themselves onto the boundary pole-skipping data. In this section, we explicitly reformulate the vacuum Einstein field equations in terms of the reconstructed metric parameters. This procedure translates the differential constraints of General Relativity into algebraic relations among the pole-skipping data. 
It is important to clarify the scope of the following analysis. While we utilized charged backgrounds (such as RN-AdS and KN-AdS) in previous sections to validate the robustness of the reconstruction formalism, the dynamical constraints derived here are based on the vacuum Einstein equations. Therefore, our discussion in this section is strictly restricted to neutral spacetimes (setting charge $Q=0$). Consequently, our consistency checks apply to Schwarzschild-AdS, rotating BTZ, and Kerr-AdS geometries, where the matter energy-momentum tensor vanishes. Subsequently, we investigate further restrictions imposed by the Null Energy Condition (NEC).

\subsection{Reinterpretation of the Einstein equations}

We focus on the vacuum Einstein equations with a negative cosmological constant,
\begin{equation}\label{EFE}
    \mathcal{E}_{\mu\nu} \equiv R_{\mu\nu} - \frac{1}{2}g_{\mu\nu}R + \Lambda g_{\mu\nu} = 0 \,,
\end{equation}
where the cosmological constant is fixed as $\Lambda = -d(d+1)/2$, setting the AdS radius to unity ($L=1$). We apply this dynamical condition individually to the black hole geometries reconstructed in the previous sections.

\subsubsection{Maximally symmetric case}
Consistent with the framework in section~\ref{Remaxbh}, we employ the static ansatz \eqref{anst1} as a solution to \eqref{EFE}. The non-vanishing components of the Einstein tensor are the temporal component $\mathcal{E}_{tt}$, the radial component $\mathcal{E}_{zz}$, and the transverse components $\mathcal{E}_{ii}$ (for $i=1,\dots,d$). Explicitly, the temporal and radial components take the form
\begin{equation}\label{eq:Ettzz_max}
\begin{split}
    \mathcal{E}_{tt} &= d(d+1)f_1 - d(d+1)f_1f_2 + d\, z f_1 f_2' + d(d-1)z^2 f_1 \kappa \,, \\[6pt]
    \mathcal{E}_{zz} &= d(d+1)f_1 - d(d+1)f_1f_2 + d\, z f_1' f_2 + d(d-1)z^2 f_1 \kappa \,.
\end{split}
\end{equation}
Here, $\kappa \in \{0, \pm 1\}$ characterizes the planar, spherical, and hyperbolic horizon topologies, respectively. While the transverse components $\mathcal{E}_{ii}$ involve complex second-order derivatives, the system is constrained by the Bianchi identities, $\nabla^\mu \mathcal{E}_{\mu\nu}=0$. Consequently, the equation $\mathcal{E}_{ii}=0$ is not independent; it is automatically satisfied provided the dynamical equation $\mathcal{E}_{tt}=0$ and the constraint equation $\mathcal{E}_{zz}=0$ hold. Therefore, it suffices to focus on the components in \eqref{eq:Ettzz_max}.

We perform a near-horizon expansion of these components in powers of $(z-1)$
\begin{equation}
\begin{split}
    \mathcal{E}_{tt} &= \mathcal{E}_{tt}^{(0)} + \mathcal{E}_{tt}^{(1)}(z-1) + \mathcal{E}_{tt}^{(2)}(z-1)^2 + \cdots \,, \\[4pt]
    \mathcal{E}_{zz} &= \mathcal{E}_{zz}^{(0)} + \mathcal{E}_{zz}^{(1)}(z-1) + \mathcal{E}_{zz}^{(2)}(z-1)^2 + \cdots \,.
\end{split}
\end{equation}
Imposing the vacuum equations requires each coefficient in these series to vanish. We observe that the zeroth-order coefficients vanish identically ($\mathcal{E}_{tt}^{(0)} = \mathcal{E}_{zz}^{(0)} = 0$). For higher orders ($n \ge 1$), the explicit expressions in terms of metric derivatives are
\begin{equation}\label{Ettzz12}
\begin{split}
    \mathcal{E}_{tt}^{(1)}&=\mathcal{E}_{zz}^{(1)} = d(d+1)f_{11} + d\, f_{11}f_{21} + d(d-1)f_{11}\,\kappa \,, \\[6pt]
    \mathcal{E}_{tt}^{(2)} &= \frac{d\, f_{12}}{2}\Big[1+d+(d-1)\kappa+f_{21}\Big] + d\, f_{11}\Big[2(d-1)\kappa - d\, f_{21} + f_{22}\Big] \,, \\[6pt]
    \mathcal{E}_{zz}^{(2)} &= \frac{d\,f_{12}}{2}\Big[1+d+(d-1)\kappa\Big] + f_{11}\Big[2d(d-1)\kappa - d(d+1)f_{21}\Big] \\[4pt]
    &\qquad\qquad + d\,f_{12}f_{21} + d\,f_{11}f_{21} + \frac{d\,f_{11}f_{22}}{2} \,.
\end{split}
\end{equation}
By substituting the reconstructed metric derivatives with mass $m=0$ obtained in section~\ref{sec:iter_static} into \eqref{Ettzz12}, we reinterpret the Einstein equations as algebraic constraints on the pole-skipping data $E_{n}(\mu^k)$. The first and second-order constraints are
\begin{equation}\label{Ettzz12ps}
\begin{split}
    \mathcal{E}_{tt}^{(1)}&=\mathcal{E}_{zz}^{(1)}= -\frac{2d\,\omega_1^2}{E_{1}(\mu)}\Big[d(1+d) + d(d-1)\kappa + 2E_{1}(\mu) \Big] = 0 \,, \\[6pt]
    \mathcal{E}_{tt}^{(2)} &= \frac{d\,\omega_1^2}{4E_{1}(\mu)^3}\Big\{ d^2 E_{2}(\mu^2) \Big[1+d+(d-1)\kappa\Big]- 40 E_{1}(\mu)^3 \\[4pt]
    &\qquad+ 4d E_{1}(\mu)^2 \Big[1 + 2 E_{2}(\mu) + 3\kappa + 3d - 5d\,\kappa + 2d^2(1+\kappa)\Big]   \\[4pt]
    &\qquad  - 2d E_{1}(\mu) \Big[5E_{2}(\mu^2) + 2d E_{2}(\mu)(d+1+(d-1)\kappa)\Big]\Big\} = 0 \,,\\[6pt]
    \mathcal{E}_{zz}^{(2)} &= \frac{d\,\omega_1^2}{4E_{1}(\mu)^3}\Big\{ d^2 E_{2}(\mu^2) \Big[1+d+(d-1)\kappa\Big]- 8(4d-1) E_{1}(\mu)^3  \\[4pt]
    &\qquad  + 4d E_{1}(\mu)^2 \Big[1 - 2 E_{2}(\mu) + 3\kappa + 3d - 5d\,\kappa + 2d^2(1+\kappa)\Big]   \\[4pt]
    &\qquad - 2d E_{1}(\mu) \Big[E_{2}(\mu^2) + 2d E_{2}(\mu)(d+1+(d-1)\kappa)\Big]\Big\} = 0 \,.
\end{split}
\end{equation}
 Solving these algebraic equations for the elementary symmetric polynomials $E_{n}(\mu^k)$ yields 
 \begin{equation}\label{E112221}
\begin{gathered}
    E_{1}(\mu) = -\frac{d}{2}\big[1+d+(d-1)\kappa\big] \,, \\[6pt]
    E_{2}(\mu) = -2(d+1)(d-\kappa-d\,\kappa) \,, \quad E_{2}(\mu^2) = d(d^2-1)(1+\kappa)(1+d-\kappa+d\,\kappa) \,.
\end{gathered}
\end{equation}
This reinterpretation extends naturally to higher orders. In general, for each order $n$, the Einstein equations provide two independent algebraic conditions that uniquely determine the parameters $E_{n}(\mu^n)$ and $E_{n}(\mu^{n-1})$.

\subsubsection{Three-dimensional rotation}
Paralleling the static analysis, we verify the vacuum Einstein equations for the three-dimensional rotating ansatz \eqref{3rBH} order by order. The non-vanishing components of the Einstein tensor are $\mathcal{E}_{tt}$, $\mathcal{E}_{zz}$, $\mathcal{E}_{t\varphi}$, and $\mathcal{E}_{\varphi\varphi}$. We perform the near-horizon expansion
\begin{equation}
    \mathcal{E}_{\mu\nu} = \mathcal{E}_{\mu\nu}^{(0)} + \mathcal{E}_{\mu\nu}^{(1)}(z-1) + \mathcal{E}_{\mu\nu}^{(2)}(z-1)^2 + \cdots \,.
\end{equation}
At zeroth order, the coefficients $\mathcal{E}_{\mu\nu}^{(0)}$ vanish identically. At first order, the radial component $\mathcal{E}_{zz}^{(1)}$ vanishes identically. For the remaining components ($\mathcal{E}_{tt}^{(1)}$, $\mathcal{E}_{t\varphi}^{(1)}$, $\mathcal{E}_{\varphi\varphi}^{(1)}$), substitution of the reconstructed metric derivatives reveals that they differ only by non-zero overall factors. Thus, they reduce to a single algebraic constraint:
\begin{equation}\label{eq:E1_3Drot}
    4 + E_{1}(k)^2 - 4E_{1}(k^2) = 0 \,.
\end{equation}

At second order, the radial component $\mathcal{E}_{zz}^{(2)}=0$ yields a concise relation
\begin{equation}
    4 E_1(k)^2 + 8 E_1(k^2) - E_2(k^2) - 4 = 0 \,.
\end{equation}
The constraints from the remaining components ($\mathcal{E}_{tt}^{(2)}, \mathcal{E}_{t\varphi}^{(2)}, \mathcal{E}_{\varphi\varphi}^{(2)}$) are algebraically involved but not independent, as dictated by the Bianchi identities. We find that the constraint from $\mathcal{E}_{\varphi\varphi}^{(2)}$ reappears as a substructure within $\mathcal{E}_{t\varphi}^{(2)}$. The independent constraints can be summarized by the vanishing of the following polynomial structures
\begin{equation}
    \begin{split}
     \mathcal{E}_{\varphi\varphi}^{(2)} \propto \;& 48 E_1(k)^2 E_1(k^2) - 12 E_1(k)^4 E_1(k^2) + 48 E_1(k^2)^2 + 12 E_1(k)^2 E_1(k^2)^2  \\[4pt]
    &+ 16 E_1(k^2)^3 - 16 E_1(k^2) E_2(k^2) + 4 E_1(k)^2 E_1(k^2) E_2(k^2) + 4 E_2(k^4)\\[4pt]
    &- 8 E_1(k) E_1(k^2) E_2(k^3)  + E_1(k)^2 E_2(k^4) + 12 E_1(k^2) E_2(k^4)=0\,,
    \end{split} 
\end{equation}
and
\begin{equation}
    \begin{split}
     \mathcal{E}^{(2)}_{t\varphi} \propto \;& \bigg[32 E_1(k) E_1(k^2)  + 32 E_1(k)^3 E_1(k^2)  - 64 E_1(k) E_1(k^2)^2  \\[4pt]
     &- 8 E_1(k) E_1(k^2) E_2(k^2) + 8 E_1(k^2) E_2(k^3) \bigg]\mathfrak{w}_1 + \mathcal{E}_{\varphi\varphi}^{(2)}\,\Omega_H = 0 \,.
    \end{split} 
\end{equation}
Component $\mathcal{E}_{tt}^{(2)}$ provides no additional information, consistent with the redundancy of the system.

\subsubsection{Four-dimensional rotation}
Finally, we apply this analysis to the four-dimensional rotating metric ansatz \eqref{Ct}. The relevant non-vanishing components of the Einstein tensor include $\mathcal{E}_{tt}$, $\mathcal{E}_{rr}$, $\mathcal{E}_{t\varphi}$, $\mathcal{E}_{\varphi\varphi}$, and $\mathcal{E}_{xx}$. Given the structure of the metric, we perform a simultaneous double series expansion around the horizon ($r=1$) and the rotation axis pole ($x=1$)
\begin{equation}
    \mathcal{E}_{\mu\nu}(r,x) = \sum_{i,j=0}^{\infty} \mathcal{E}_{\mu\nu}^{(i,j)} \, (r-1)^i (x-1)^j \,.
\end{equation}
Imposing the vacuum Einstein equations $\mathcal{E}_{\mu\nu}=0$ requires each expansion coefficient $\mathcal{E}_{\mu\nu}^{(i,j)}$ to vanish independently.

At zeroth order ($i=j=0$), we find that the diagonal components $\mathcal{E}_{rr}^{(0,0)}$ and $\mathcal{E}_{xx}^{(0,0)}$ provide the leading constraints. Explicitly, they take the form
\begin{equation}\label{E4D00}
\begin{split}
    \mathcal{E}_{rr}^{(0,0)} :\quad &6a\Omega_H(\mathfrak{w}_1+k_1\Omega_H) - \mathfrak{w}_1\Omega_H\Big(E_{2}^{(r)}(\lambda) - 6E_{1}^{(r)}(\lambda)\Big) \\[4pt]
    &\qquad\qquad - k_1\Omega_H^2\Big(E_{2}^{(r)}(\lambda) - 8E_{1}^{(r)}(\lambda) + 2E_{1}^{(a)}(\lambda)\Big)  = 0 \,, \\[10pt]
    \mathcal{E}_{xx}^{(0,0)} :\quad& 6a\Omega_H(\mathfrak{w}_1+k_1\Omega_H) + k_1\Omega_H^2\Big(E_{2}^{(a)}(\lambda) - 6E_{1}^{(a)}(\lambda)\Big) \\[4pt]
    &\qquad \qquad + \mathfrak{w}_1\Omega_H\Big(E_{2}^{(a)}(\lambda) + 2E_{1}^{(r)}(\lambda) - 8E_{1}^{(a)}(\lambda)\Big) = 0 \,.
\end{split}
\end{equation}
The vanishing of these coefficients imposes strict algebraic relations among the pole-skipping parameters $E_{n}^{(r)}(\lambda^m)$ and $E_{n}^{(a)}(\lambda^m)$, serving as consistency checks for the reconstructed metric. Proceeding to higher orders in the double expansion ($i, j > 0$), the system yields a hierarchy of algebraic equations involving $\mathcal{E}_{\mu\nu}^{(i,j)}$. While the algebraic complexity increases significantly compared to the static or 3D cases, the fundamental logic remains unchanged: the vacuum Einstein equations serve as recurrence relations that constrain the higher-order pole-skipping parameters. This confirms that the metric reconstruction procedure is dynamically self-consistent, with the bulk geometry being determined by these data order by order.

\subsection{Null Energy Condition}
As demonstrated in the previous section, the bulk metric and vacuum Einstein equations can be reformulated in terms of the pole-skipping parameters. A natural consequence is that the NEC implies nontrivial inequalities for these data. Moreover, in quantum field theories where pole skipping is observed but a dual gravitational description remains unknown, analyzing these constraints may provide insights into the necessary conditions for a valid gravitational dual.

The NEC requires that
    $T_{\mu\nu}\,k^\mu k^\nu \;\ge\; 0$ for all null vectors $k^\mu$,
ensuring that the energy flux along any lightlike direction is non-negative.
In holographic setups, the NEC plays a central role in constraining admissible bulk geometries through Einstein’s equations
\begin{equation}\label{EEe}
    G_{\mu\nu} + \Lambda g_{\mu\nu} = 8\pi G_N T_{\mu\nu}\,,
\end{equation}
and serves as the foundation for various bulk no-go theorems~\cite{Freedman:1999gp, Gao:2000ga}. These geometric constraints are crucial for maintaining causality and unitarity in the dual field theory, as they rule out pathological bulk configurations that would otherwise lead to inconsistencies on the boundary.

For concreteness, we consider the metric ansatz \eqref{anst1} with $\kappa=0$ (here $\kappa$ denotes the horizon topology) and focus on a radial null vector in the \((t,z)\) plane
\[
k^\mu = \bigl(1,\;\sqrt{f_1(z)\,f_2(z)},\;0,\dots,0\bigr)\,,
\]
which satisfies \(g_{\mu\nu}k^\mu k^\nu = 0\) by construction. Under Einstein’s equations \eqref{EEe}, the NEC reduces to the condition
\begin{equation}
    \mathcal{F}(z) \equiv G_{\mu\nu}\,k^\mu k^\nu = f_1(z)\,f_2'(z) - f_2(z)\,f_1'(z) \ge 0\,.
\end{equation}
We now examine this inequality near the black hole horizon at \(z = z_h\). Expanding \( \mathcal{F}(z) \) around the horizon, we find that \( \mathcal{F}(z_h) = \mathcal{F}'(z_h) = 0 \). Hence, the leading nontrivial contribution arises at the second order, and the NEC implies $\mathcal{F}''(z_h) \ge 0$. Evaluating this derivative yields
\begin{equation}\label{nec}
    f_{11}\,f_{22} - f_{21}\,f_{12} \ge 0\,.
\end{equation}

This yields a concrete constraint on the near-horizon expansion coefficients of the metric functions.
Substituting the reconstructed near-horizon derivatives of the metric functions from \eqref{f11}, \eqref{f21}, and \eqref{f212} into \eqref{nec}, we obtain the following constraint relating the momentum squared $\mu$ among different pole-skipping points
\begin{equation}\label{uct}
    4\mu_{11}^2 + d(2\mu_{11} - \mu_{21})(2\mu_{11} - \mu_{22}) \ge 0\,.
\end{equation}

In addition to the NEC constraint \eqref{uct}, the metric must admit a regular causal structure.
 To ensure a standard causal structure where $t$ remains timelike  $(g_{tt}<0)$ just outside the horizon ($z < z_h$), we require $f_{11}<0$. Using the relation from \eqref{f11}, this translates directly to the constraint $\mu_{11}<0$.

\section{Algebraic constraints on pole-skipping data}
\label{sec:universality}
Previous studies on static planar black holes have established that pole-skipping points are not independent; rather, they are governed by intrinsic homogeneous polynomial identities~\cite{Lu:2025pal, Lu:2025jgk}. Here, we investigate this observation in a more general scenario and demonstrate that these constraints fundamentally originate from the overdetermined nature of the bulk reconstruction problem. 

This overdetermined structure stems from a specific algebraic feature of the near-horizon (or near-axis) expansion: the highest-order metric derivatives at level $n$ appear exclusively and linearly within the coefficients of the lowest powers of the pole-skipping polynomial. This linearity is the pivotal mathematical property that guarantees the invertibility of the recursive reconstruction equations.

Let $\mathcal{N}$ denote the degree of the determinant polynomial (corresponding to the number of pole-skipping roots at a given order) and $\mathfrak{q}$ denote the number of coupled metric functions (the unknowns).
To systematically conceptualize how this overdetermined mechanism ($\mathcal{N} > \mathfrak{q}$) manifests across different spacetime geometries, we summarize the structural distinctions in table~\ref{tab:universality_classes}. While the specific values of $\mathcal{N}$ and $\mathfrak{q}$ vary depending on the spacetime symmetries and the separability of the wave equation, the underlying mathematical principle remains invariant: this redundant algebraic structure naturally restricts the allowable phase space of the boundary theory, yielding exactly $\mathcal{N} - \mathfrak{q}$ independent algebraic constraints on the pole-skipping data.

\begin{table}[htbp]
\centering
\vspace{2mm}
\renewcommand{\arraystretch}{2} 
\resizebox{\textwidth}{!}{
\begin{tabular}{@{}lccc@{}}
\toprule
\textbf{Spacetime Geometry} & \textbf{Static Case} & \textbf{3D Rotating} & \textbf{4D Rotating (Separable)} \\ \midrule
\textbf{Symmetry} & Maximally symmetric & Axisymmetric & Axisymmetric + Hidden Symmetry \\
\textbf{Metric Unknowns ($\mathfrak{q}$)} &  ($f_1, f_2$) &  ($f_1, f_2, f_3$) &  $f_{1,2}$ for radial, $y_{1,2}$ for angular \\
\textbf{Polynomial Variable} & $\mu$ & $k$ & $\lambda$ (Separation Constant) \\
\textbf{Polynomial Degree ($\mathcal{N}$)} & $n$ & $2n$ & $n$ (Radial) + $n$ (Angular) \\
\textbf{Determinant Source} & Near-horizon & Near-horizon & Near-horizon (Radial) + Near-axis (Angular) \\
\textbf{Constraint Count ($\mathcal{N} - \mathfrak{q}$)} & $n - 2$ & $2n - 3$ & $n - 2$ (per independent sector) \\ \bottomrule
\end{tabular}
}
\caption{Structural comparison of the pole-skipping reconstruction framework across different black hole geometries. The parameters $\mathcal{N}$ and $\mathfrak{q}$ denote the degree of the determinant polynomial (number of roots) and the number of coupled metric unknowns, respectively. The intrinsically overdetermined nature of the near-horizon and near-axis expansions ($\mathcal{N} > \mathfrak{q}$) systematically yields $\mathcal{N} - \mathfrak{q}$ independent algebraic constraints on the boundary data.}
\label{tab:universality_classes}
\end{table}

This structure necessitates a separation in the role of the expansion coefficients derived from Vieta's formulas:
\begin{itemize}
\item \textbf{Reconstruction sector:} The $\mathfrak{q}$ equations associated with the lowest powers of the polynomial (indices $0, \dots, \mathfrak{q}-1$), which explicitly contain the linear $n$th-order metric derivative terms (see e.g., \eqref{VF1}), serve as the reconstruction equations that uniquely determine the metric coefficients.
\item \textbf{Redundancy sector:} The remaining $\mathcal{N} - \mathfrak{q}$ equations involve polynomial coefficients $V_{\mathcal{N},j}$ (with $j \ge \mathfrak{q}$) that do not depend on the $n$th-order derivatives. Instead, they serve as intrinsic polynomial constraints on the pole-skipping data itself. These constraints can be written generally via Vieta’s formulas as
\begin{equation}\label{Consec}
P_{n}(\zeta^m) \equiv (-1)^{m}V_{n,\,\mathcal{N}}\,E_{n}(\zeta^m)-V_{n,\,\mathcal{N}-m}=0\,,\quad m=1,\cdots,\mathcal{N}-\mathfrak{q}\,,
 \end{equation}
where $\zeta$ denotes the pole-skipping parameter: $\zeta=\mu$ for the maximally symmetric case, $\zeta=k$ for three-dimensional rotation, and $\zeta=\lambda$ for four-dimensional rotation. $E_{n}(\zeta^m)$ denotes the elementary symmetric polynomial of degree $m$ (consistent with \eqref{ek1}).
\end{itemize}

Physically, these polynomial constraints imply that the pole-skipping points cannot be arbitrarily distributed in the complex momentum plane. For a boundary theory to admit a dual classical gravitational description, its pole-skipping data should satisfy these redundant algebraic relations.

\subsection{\texorpdfstring{$\mu$-- and $\lambda$--polynomial constraints}{mu- and lambda-polynomial constraints}}

We first consider the class of problems characterized by $\mathcal{N}=n$ and $\mathfrak{q}=2$. This universality class includes both the static maximally symmetric black holes and the separable sectors of four-dimensional rotating black holes. In these cases, for any order $n \ge 3$, the system is over-determined, yielding $n-2$ algebraic constraints.

\subsubsection{\texorpdfstring{$\mu$--polynomial constraints}{mu-polynomial constraints}}\label{muc}

Applying the consistency condition to the static case, we recover the results of~\cite{Lu:2025pal, Lu:2025jgk}.
At the third order ($n=3$), there is exactly one redundant equation ($3-2=1$). This constraint arises from the coefficient of the $\mu^2$ term in the determinant polynomial, which is independent of third-order metric derivatives. The condition requires
\begin{equation}
    P_3(\mu) \equiv -V_{3,3}\, E_{3}(\mu) - V_{3,2} = 0\,.
\end{equation}
Substituting the reconstructed lower-order metric coefficients from \eqref{f11} and \eqref{f212} into $V_{3,2}$ and $V_{3,3}$, we derive the explicit constraint\footnote{Hereafter, we drop overall non-zero factors in the expressions of $P$.}
\begin{equation}\label{eq:mu_constraint_n3}
   P_3(\mu) = E_3(\mu) - 4 E_2(\mu) + 5 E_1(\mu) = 0 \,.
\end{equation}
At the fourth order ($n=4$), there are $4-2=2$ consistency constraints, corresponding to the coefficients of $\mu^3$ and $\mu^2$
\begin{equation}\label{V4_rev}
\begin{split}
    P_{4}(\mu) &\equiv - V_{4,4}\, E_{4}(\mu) - V_{4,3} = 0 \,, \\[6pt]
    P_{4}(\mu^2) &\equiv V_{4,4}\, E_{4}(\mu^2) - V_{4,2} = 0 \,.
\end{split}
\end{equation}
Using the reconstructed metric derivatives $f_{ai}$ ($i<4$), these yield the specific algebraic constraints
\begin{equation}\label{E423_rev}
\begin{split}
   P_{4}(\mu) &= E_4(\mu) - 10 E_2(\mu) + 16 E_1(\mu) = 0 \,, \\[6pt]
   P_{4}(\mu^2) &= E_4(\mu^2) - 6 E_3(\mu^2) + 14 E_2(\mu^2) - 9 E_2(\mu)^2 + 40 E_2(\mu)\,E_1(\mu) - 46 E_1(\mu)^2 = 0 \,.
\end{split}
\end{equation}
The general recursive formula for $P_n(\mu^m)$ for arbitrary $n$ can be found in~\cite{Lu:2025pal, Lu:2025jgk}.

\subsubsection{\texorpdfstring{$\lambda$--polynomial constraints}{lambda-polynomial constraints}}

For four-dimensional rotating black holes discussed in section~\ref{sec:recon_4d}, the reconstruction decouples into independent radial and angular sectors. Both sectors are governed by degree-$n$ polynomials in the separation constant $\lambda$, with two unknowns per sector ($f_{1n}, f_{2n}$ for radial; $y_{1n}, y_{2n}$ for angular).

Mathematically, this structure is strictly isomorphic to the static case. The near-horizon (or near-pole) Frobenius recurrence relations dictate the algebraic form of the determinant, independent of the specific values of the background parameters (such as mass, charge, or rotation). Consequently, the degree of over-determination is $\mathcal{N} - \mathfrak{q} = n - 2$, and the eigenvalues $\bm{\lambda}^{(r)}_n$ and $\bm{\lambda}^{(a)}_n$ must satisfy the  same hierarchy of constraints as \eqref{eq:mu_constraint_n3} and \eqref{E423_rev}.

For explicit demonstration, at $n=3$, the radial (and angular) data must satisfy
\begin{equation}
     P_3^{s}(\lambda)= E^{(s)}_3(\lambda) - 4 E^{(s)}_2(\lambda) + 5 E^{(s)}_1(\lambda) = 0 \,, \quad \text{for}\, s \in \{r, a\} \,.
\end{equation}
Given this structural isomorphism, the general form of the constraint polynomials $P_n^{s}(\lambda^m)$ for arbitrary $n$ is identical to the static results presented in section~\ref{muc}.
This confirms that the algebraic constraints are universal to the class of differential operators with $\mathfrak{q}=2$ coupled functions, transcending the distinction between static and rotating spacetimes.
\subsection{\texorpdfstring{$k$--polynomial constraints}{k-polynomial constraints}}
In stark contrast to the static and separable four-dimensional cases, the three-dimensional rotating black hole exhibits a distinct algebraic structure. 
Specifically, the determinant condition yields a polynomial in $k$ of degree $\mathcal{N}=2n$, whereas the reconstruction is governed by only $\mathfrak{q}=3$ coupled metric functions ($f_{1n}, f_{2n}, f_{3n}$). This structural asymmetry leads to a higher degree of over-determination compared to the static case:
\begin{equation}
    N_{\text{constraints}} = \mathcal{N} - \mathfrak{q} = 2n - 3 \quad (\text{for } n \ge 2)\,.
\end{equation}
Physically, these constraints imply that the pole-skipping momenta $k_n$ cannot be arbitrarily distributed in the complex plane; they must satisfy intrinsic algebraic conditions to be compatible with a geometric dual described by a stationary axisymmetric ansatz.

At the lowest order ($n=1$), the pole-skipping data, combined with the Hawking temperature, exactly determines the metric derivatives, leaving no redundant information. However, for all $n \ge 2$, the system becomes strictly overdetermined, yielding intrinsic algebraic constraints.

At order $n=2$, the polynomial degree is $\mathcal{N}=4$, while there are $\mathfrak{q}=3$ unknowns. This yields exactly $4-3=1$ constraint. The reconstruction utilizes the coefficients of $k^0, k^1, k^2$ to determine the metric derivatives. The constraint arises from the coefficient of $k^3$, which must satisfy the relation dictated by Vieta's formulas. This imposes the following condition on the elementary symmetric polynomials of the wavenumbers
\begin{equation}
    P_2(k) = E_2(k)-4 E_1(k) = 0 \,.
\end{equation}

At order $n=3$, the number of constraints increases to $2(3)-3=3$. The pole-skipping data must satisfy the hierarchy
\begin{equation}
    \begin{split}
        P_3(k) &= E_3(k)-9 E_1(k)=0\,, \\[5pt]
        P_3(k^2) &= E_3(k^2)- 4 E_2(k^2)-11 E_1(k)^2 + 5 E_1(k^2)=0\,, \\[5pt]
        P_3(k^3) &= E_3(k^3) - 6 E_2(k^3) - 12 E_1(k)\, E_2(k^2) + 30 E_1(k)\, E_1(k^2) + 21 E_1(k)^3=0 \,.
    \end{split}
\end{equation}

At order $n=4$, the system becomes highly constrained with $2(4)-3=5$ independent algebraic relations. The first three constraints follow the pattern of lower orders
\begin{equation}
    \begin{split}
        P_4(k) &=  E_4(k)-16 E_1(k)=0\,, \\[5pt]
        P_4(k^2) &= E_4(k^2) - 10 E_2(k^2)-56 E_1(k)^2 + 16 E_1(k^2)=0\,, \\[5pt]
        P_4(k^3) &= E_4(k^3) - 20 E_2(k^3)  + 192 E_1(k)\, E_1(k^2) - 80 E_1(k) E_2(k^2)+64 E_1(k)^3=0\,.
    \end{split}
\end{equation}
The remaining two highest-order constraints correspond to the coefficients of $k^4$ and $k^5$:
\begin{equation}
    \begin{split}
        P_4(k^4) &= E_4(k^4) - 6 E_3(k^4)+ 40 E_1(k^2)\, E_2(k^2) - 9 E_2(k^2)^2 - 52 E_1(k)\, E_2(k^3) \\
        &\quad + 14 E_2(k^4) + 240 E_1(k)^4+ 338 E_1(k)^2\, E_1(k^2) \\
        &\quad - 46 E_1(k^2)^2 - 88 E_1(k)^2\, E_2(k^2)=0\,,
    \end{split}
\end{equation}
and
\begin{equation}
    \begin{split}
        P_4(k^5) &= E_4(k^5)- 8 E_3(k^5)-696 E_1(k)^3\, E_1(k^2) - 368 E_1(k)\, E_1(k^2)^2 \\
        &\quad + 160 E_1(k)\, E_1(k^2)\, E_2(k^2)+ 256 E_1(k)^2\, E_2(k^3) + 80 E_1(k^2)\, E_2(k^3) \\
        &\quad - 36 E_2(k^2)\, E_2(k^3) + 112 E_1(k)\, E_2(k^4) - 24 E_1(k)\, E_3(k^4)=0 \,.
    \end{split}
\end{equation}

Although deriving a closed-form expression for the full set of constraints for arbitrary $n$ is analytically intractable, a universal pattern emerges for the leading constraint $P_n(k)$. Based on the calculated results for $n=2, \dots, 5$, we infer that the linear symmetric polynomials satisfy the universal relation
\begin{equation}\label{pnk}
    P_n(k) = E_n(k) - n^2 E_1(k) = 0 \quad (\text{for } n \ge 2)\,.
\end{equation}
The linear growth of the number of constraints ($2n-3$) highlights the restrictive nature of the bulk geometry on the boundary data. 

\section{Discussion}
\label{sec:discussion}
In this work, we have established a robust analytic framework for reconstructing the bulk geometry of stationary, axisymmetric spacetimes using pole-skipping data. By extending the reconstruction scheme from maximally symmetric black holes to rotating systems, we have demonstrated that the near-horizon metric derivatives are encoded within the algebraic structure of the probe wave equation’s characteristic polynomials.

Ultimately, this work addresses a critical question regarding the universality of bulk reconstruction: Does it rely on the maximal spacetime symmetry? Our analysis reveals that the reconstruction scheme extends well beyond such symmetric constraints. For four-dimensional rotating black holes, the separability of the KG equation is a pivotal property that allows the coupled partial differential equations to decouple. Crucially, we demonstrated that relying solely on conventional pole-skipping data derived from near-horizon expansion is insufficient to reconstruct the complete bulk geometry. However, by introducing angular pole-skipping via near-axis analysis, we obtain a complementary dataset. These two distinct sets of data—radial and angular—independently determine their corresponding metric functions, thereby achieving a full reconstruction of the spacetime throughout its analytic domain despite the intricacies of rotation.

The reinterpretation of the vacuum Einstein equations as algebraic relations among pole-skipping points provides a consistency check for the proposed holographic dual. Furthermore, the algebraic inequalities involving pole-skipping parameters derived from the NEC demonstrate that the physical viability of a bulk geometry is directly imprinted on its boundary observables.

A particularly salient feature of our framework is the overdetermined nature of the reconstruction problem. Since the number of pole-skipping roots $\mathcal{N}$ typically exceeds the number of coupled metric functions $\mathfrak{q}$, the excess information manifests as general polynomial constraints on the boundary data. 
These conditions, such as the relation \eqref{pnk} obtained via induction for three-dimensional rotation, reveal that pole-skipping points are rigidly constrained in the complex momentum plane. Physically, these constraints serve as a geometric fingerprint, reflecting the highly overdetermined mapping from boundary observables to the bulk spacetime.

We outline several avenues for future exploration:
\begin{itemize}
    \item \textbf{Holographic origin of angular pole-skipping}: Although the bulk description of angular pole-skipping is mathematically well-defined through near-axis analysis, its precise interpretation in the dual field theory remains to be elucidated.
    \item \textbf{Non-separable backgrounds}: A natural next step is to address spacetimes that do not admit a separable coordinate system. Such cases would require solving the "reconstruction sector" for partial differential equations, possibly necessitating new techniques for extracting pole-skipping conditions from coupled systems.
    \item \textbf{Backreaction and beyond scalar probes}: Extending the reconstruction scheme to include the backreaction of the probe field or considering spinor, vector, and gravitational perturbations may reveal even more intricate algebraic constraints on the pole-skipping data.
    \item \textbf{Higher-dimensional rotating spacetimes}: Our reconstruction scheme may also extend to rotating spacetimes with a separable KG equation in higher dimensions~\cite{Gibbons:2004js, Gibbons:2004uw, Chen:2006xh, Kolar:2015hea}. A distinct feature is that the number of decoupled angular differential equations and unknown metric functions increases in higher dimensions. This necessitates the introduction of angular pole-skipping for each angular sector to reconstruct the corresponding angular metric. Another feature is that pole skipping relies on multiple parameters, including Carter constants and azimuthal quantum numbers, and forms a hypersurface in parameter space. It remains to be investigated how to select an economical and systematic subset of the data for the reconstruction. Additionally, whether the linearity of the pole-skipping polynomial with respect to the highest-order metric derivatives holds universally remains unknown.
\end{itemize}

\acknowledgments
We would like to thank Xian-Hui Ge, Yan Liu and Zhuo-Yu Xian
for their helpful discussions. SFW is supported by NSFC grants No.12275166 and No.12311540141.

\appendix
\section{Explicit expressions for the elementary symmetric polynomials}
\subsection{The RN-AdS black hole}\label{data:RN}
We provide the explicit closed-form expressions for the elementary symmetric polynomials $E_n(\mu^k)$ associated with the Reissner-Nordstr\"om-AdS$_4$ black hole, evaluated up to the fifth order ($n=5$). These expressions serve as the pole-skipping data utilized in section~\ref{sec:RN_example} for the explicit metric reconstruction.

At the first order ($n=1$), there is only one term:
\begin{equation}
    E_1(\mu) = 4\pi T \,.
\end{equation}

At the second order ($n=2$), the symmetric polynomials are given by:
\begin{equation}
\begin{split}
    E_2(\mu) &= 4(3 + 8\pi T) \,, \\
    E_2(\mu^2) &= 32\pi T(3 + 5\pi T) \,.
\end{split}
\end{equation}

At the third order ($n=3$), the expressions are:
\begin{equation}
\begin{split}
    E_3(\mu) &= 12(4 + 9\pi T) \,, \\
    E_3(\mu^2) &= 48(9 + 52\pi T + 63\pi^2 T^2) \,, \\
    E_3(\mu^3) &= 576\pi T(9 + 34\pi T + 31\pi^2 T^2) \,.
\end{split}
\end{equation}

At the fourth order ($n=4$), we have:
\begin{equation}
\begin{split}
    E_4(\mu) &= 8(15 + 32\pi T) \,, \\
    E_4(\mu^2) &= 48(81 + 388\pi T + 432\pi^2 T^2) \,, \\
    E_4(\mu^3) &= 128(243 + 2232\pi T + 5790\pi^2 T^2 + 4544\pi^3 T^3) \,, \\
    E_4(\mu^4) &= 2048\pi T(243 + 1503\pi T + 2982\pi^2 T^2 + 1910\pi^3 T^3) \,.
\end{split}
\end{equation}
Finally, at the fifth order ($n=5$), the expressions expand to:
\begin{equation}
\begin{split}
    E_5(\mu) &= 20(12 + 25\pi T) \,, \\
    E_5(\mu^2) &= 16(1143 + 5124\pi T + 5500\pi^2 T^2) \,, \\
    E_5(\mu^3) &= 64(7776 + 58707\pi T + 137646\pi^2 T^2 + 102100\pi^3 T^3) \,, \\
    E_5(\mu^4) &= 1280(2916 + 37368\pi T + 153333\pi^2 T^2 + 254280\pi^3 T^3 + 148015\pi^4 T^4) \,, \\
    E_5(\mu^5) &= 25600\pi T(2916 + 25704\pi T + 81621\pi^2 T^2 + 111378\pi^3 T^3 + 55375\pi^4 T^4) \,.
\end{split}
\end{equation}

\subsection{The Kerr-Newman-AdS black hole}\label{data:KN}

We present the explicit closed-form expressions for the elementary symmetric polynomials $E^{(r)}_n(\lambda^k)$ and $E^{(a)}_n(\lambda^k)$ in section~\ref{sec:KN_example}, which correspond to the radial and angular sectors of the Kerr-Newman-AdS$_4$ black hole, respectively. Due to the coupling introduced by rotation, these expressions are significantly more intricate than those of the static case.

\subsubsection*{Radial Sector Data $E^{(r)}_n(\lambda^k)$}

At the first order ($n=1$), we have:
\begin{equation}
    E^{(r)}_1(\lambda) = -\frac{2 i a (a^2-1)}{1+a^2} - 4 \pi T \,.
\end{equation}
At the second order ($n=2$):
\begin{equation}
\begin{split}
    E^{(r)}_2(\lambda) &= 2 \bigg[ 7 + a^2 - \frac{8 i a (a^2-1)}{1+a^2} - 16 \pi T \bigg] \,, \\
    E^{(r)}_2(\lambda^2) &= \frac{8}{(1+a^2)^2} \bigg[ i a^3 (1 + 2 \pi T) - i a^7 (1 + 2 \pi T) + 2 \pi T (10 \pi T - 7) \\
    & \quad - i a (22 \pi T - 7) + i a^5 (22 \pi T - 7) + 6 a^4 (2 - 3 \pi T + 2 \pi^2 T^2) \\
    & \quad - 2 a^6 (3 + \pi T + 2 \pi^2 T^2) + 6 a^2 (6 \pi^2 T^2 - 5 \pi T - 1) \bigg] \,.
\end{split}
\end{equation}
At the third order ($n=3$):
\begin{equation}
\begin{split}
    E^{(r)}_3(\lambda) &= 56 + 8 a^2 - \frac{54 i a (a^2-1)}{1+a^2} - 108 \pi T \,, \\
    E^{(r)}_3(\lambda^2) &= \frac{12}{(1+a^2)^2} \bigg[ 49 + a^8 - 244 \pi T + 252 \pi^2 T^2 + 6 i a^3 (3 + 2 \pi T) \\
    & \quad - 6 i a^7 (3 + 2 \pi T) - 6 i a (44 \pi T - 21) + 6 i a^5 (44 \pi T - 21) \\
    & \quad + 12 a^4 (18 - 25 \pi T + 17 \pi^2 T^2) - a^6 (53 + 28 \pi T + 24 \pi^2 T^2) \\
    & \quad + a^2 (43 - 516 \pi T + 480 \pi^2 T^2) \bigg] \,, 
    \end{split}
\end{equation}
\begin{equation}
\begin{split}
    E^{(r)}_3(\lambda^3) &= \frac{72}{(1+a^2)^3} \bigg[ -i a^{11} (1 + 4 \pi T) - 2 a^{10} (7 + 15 \pi T + 4 \pi^2 T^2) \\
    & \quad + 2 i a^9 (15 + 6 \pi T + 56 \pi^2 T^2) - 2 \pi T (49 - 160 \pi T + 124 \pi^2 T^2) \\
    & \quad + 2 i a^3 (9 - 186 \pi T + 158 \pi^2 T^2) - i a^7 (197 - 376 \pi T + 316 \pi^2 T^2) \\
    & \quad + i a (49 - 356 \pi T + 428 \pi^2 T^2) - i a^5 (540 \pi^2 T^2 - 344 \pi T - 101) \\
    & \quad + a^4 (84 - 650 \pi T + 880 \pi^2 T^2 - 408 \pi^3 T^3) \\
    & \quad + 4 a^8 (28 \pi^3 T^3 - 12 \pi^2 T^2 + 59 \pi T - 21) \\
    & \quad + 2 a^6 (56 - 201 \pi T + 112 \pi^2 T^2 + 44 \pi^3 T^3) \\
    & \quad - 2 a^2 (49 + 40 \pi T - 468 \pi^2 T^2 + 316 \pi^3 T^3) \bigg] \,.
\end{split}
\end{equation}
At the fourth order ($n=4$), the expressions become significantly more involved. The first two terms are:
\begin{equation}
\begin{split}
    E^{(r)}_4(\lambda) &= 4 \bigg[ 35 + 5 a^2 - \frac{32 i a (a^2-1)}{1+a^2} - 64 \pi T \bigg] \,, \\
    E^{(r)}_4(\lambda^2) &= \frac{4}{(1+a^2)^2} \bigg[ 1323 + 27 a^8 - 5456 \pi T + 5184 \pi^2 T^2 + 80 i a^3 (5 + 2 \pi T) \\
    & \quad - 80 i a^7 (5 + 2 \pi T) - 16 i a (334 \pi T - 175) + 16 i a^5 (334 \pi T - 175) \\
    & \quad - 16 a^6 (59 + 41 \pi T + 20 \pi^2 T^2) + 16 a^2 (103 - 723 \pi T + 628 \pi^2 T^2) \\
    & \quad + a^4 (4858 - 6768 \pi T + 4544 \pi^2 T^2) \bigg] \,,
\end{split}
\end{equation}
while the higher powers are given by
\begin{equation}
\begin{split}
    E^{(r)}_4(\lambda^3) &= \frac{16}{(1+a^2)^3} \bigg[ 3087 + 9 a^{12} - 24528 \pi T + 54544 \pi^2 T^2 - 36352 \pi^3 T^3 \\
    & \quad - 24 i a^{11} (11 + 14 \pi T) - 8 a^{10} (251 + 314 \pi T + 78 \pi^2 T^2) \\
    & \quad - 112 i a^7 (297 - 566 \pi T + 448 \pi^2 T^2) \\
    & \quad + 8 i a^9 (209 + 542 \pi T + 1088 \pi^2 T^2) \\
    & \quad - 48 i a^5 (1408 \pi^2 T^2 - 1126 \pi T - 165) \\
    & \quad + 8 i a^3 (1375 - 7882 \pi T + 6272 \pi^2 T^2) \\
    & \quad + 8 i a (1617 - 7298 \pi T + 7360 \pi^2 T^2) \\
    & \quad - 16 a^6 (640 \pi^3 T^3 - 3394 \pi^2 T^2 + 4534 \pi T - 1589) \\
    & \quad + a^8 (8704 \pi^3 T^3 - 1328 \pi^2 T^2 + 26928 \pi T - 11427) \\
    & \quad - 8 a^2 (623 + 5978 \pi T - 20522 \pi^2 T^2 + 12544 \pi^3 T^3) \\
    & \quad - 3 a^4 (27648 \pi^3 T^3 - 54880 \pi^2 T^2 + 41760 \pi T - 8921) \bigg] \,,
\end{split}
\end{equation}
\begin{equation}
\begin{split}
    E^{(r)}_4(\lambda^4) &= \frac{128}{(1+a^2)^4} \bigg[ -9 i a^{15} (1 + 6 \pi T) - 6 a^{14} (35 + 143 \pi T + 54 \pi^2 T^2) \notag\\
    & \quad + i a^{13} (1297 + 2738 \pi T + 3024 \pi^2 T^2 + 864 \pi^3 T^3) \notag\\
    & \quad + i a^9 (3552 \pi^3 T^3 - 71856 \pi^2 T^2 + 98666 \pi T - 26771) \notag\\
    & \quad + 2 \pi T (15280 \pi^3 T^3 - 28192 \pi^2 T^2 + 16590 \pi T - 3087) \notag\\
    & \quad - 3 i a (24096 \pi^3 T^3 - 32720 \pi^2 T^2 + 12390 \pi T - 1029) \notag\\
    & \quad - i a^{11} (32640 \pi^3 T^3 - 19200 \pi^2 T^2 + 23390 \pi T - 5819) \notag\\
    & \quad + i a^5 (67872 \pi^3 T^3 - 29328 \pi^2 T^2 - 64234 \pi T + 22387) \notag\\
    & \quad - i a^3 (110208 \pi^3 T^3 - 183168 \pi^2 T^2 + 57290 \pi T + 3031) \notag\\
    & \quad + i a^7 (142848 \pi^3 T^3 - 202368 \pi^2 T^2 + 80734 \pi T - 2779) \notag\\
    & \quad + 6 a^{12} (70 + 101 \pi T + 2670 \pi^2 T^2 + 448 \pi^3 T^3 + 144 \pi^4 T^4)\notag \\
    & \quad - 2 a^6 (4160 \pi^4 T^4 + 51904 \pi^3 T^3 - 132330 \pi^2 T^2 + 77511 \pi T - 3465) \notag\\
    & \quad - 2 a^{10} (9440 \pi^4 T^4 - 14624 \pi^3 T^3 + 32898 \pi^2 T^2 - 26727 \pi T + 11655) \notag\\
    & \quad - 2 a^8 (26800 \pi^4 T^4 - 16864 \pi^3 T^3 + 966 \pi^2 T^2 + 10605 \pi T - 13020) \notag\\
    & \quad + 2 a^2 (49952 \pi^4 T^4 - 103520 \pi^3 T^3 + 35898 \pi^2 T^2 + 14349 \pi T - 5145) \notag\\
    & \quad + a^4 (94880 \pi^4 T^4 - 261632 \pi^3 T^3 + 223068 \pi^2 T^2 - 46950 \pi T + 420) \bigg] \,.
\end{split}
\end{equation}
For brevity, we omit the fully expanded $n \ge 5$ terms of the radial sector, as they follow the identical structure and can be systematically generated using the Vieta's relations provided in the main text.

\subsubsection*{Angular Sector Data $E^{(a)}_n(\lambda^k)$}

The construction of the angular sector polynomials follows a similar decomposition. The lower orders read:
\begin{equation}
\begin{split}
    E^{(a)}_1(\lambda) &= \frac{2 a (a^3 - a - 2 i \pi T - 2 i a^2 \pi T)}{1+a^2} \,, \\
    E^{(a)}_2(\lambda) &= \frac{2 - 24 a^2 + 6 a^4 - 32 i a \pi T - 32 i a^3 \pi T}{1+a^2} \,, \\
    E^{(a)}_2(\lambda^2) &= \frac{16 a}{(1+a^2)^2} \bigg[ -2 i \pi T + 14 i a^2 \pi T + 10 i a^4 \pi T - 6 i a^6 \pi T \\
    & \quad + a^3 (6 - 24 \pi^2 T^2) - a (1 + 12 \pi^2 T^2) - a^5 (5 + 12 \pi^2 T^2) \bigg] \,.
\end{split}
\end{equation}

At the third order ($n=3$):
\begin{equation}
\begin{split}
    E^{(a)}_3(\lambda) &= \frac{2 (4 - 43 a^2 + 7 a^4 - 54 i a \pi T - 54 i a^3 \pi T)}{1+a^2} \,, \\
    E^{(a)}_3(\lambda^2) &= \frac{12}{(1+a^2)^2} \bigg[ 1 + 2 a^8 - 48 i a \pi T + 372 i a^3 \pi T + 336 i a^5 \pi T \\
    & \quad - 84 i a^7 \pi T + a^4 (167 - 552 \pi^2 T^2) - 12 a^2 (3 + 23 \pi^2 T^2) \\
    & \quad - 2 a^6 (35 + 138 \pi^2 T^2) \bigg] \,, \\
    E^{(a)}_3(\lambda^3) &= \frac{72 a}{(1+a^2)^3} \bigg[ -6 i \pi T - 12 i a^{10} \pi T + 24 i a^8 \pi T (13 + 15 \pi^2 T^2) \\
    & \quad - 2 a^9 (5 + 102 \pi^2 T^2) - a (3 + 112 \pi^2 T^2) + 2 i a^2 \pi T (73 + 180 \pi^2 T^2) \\
    & \quad + a^7 (129 + 356 \pi^2 T^2) + 2 i a^4 \pi T (540 \pi^2 T^2 - 217) \\
    & \quad + 2 i a^6 \pi T (540 \pi^2 T^2 - 131) + a^3 (51 + 540 \pi^2 T^2) + a^5 (1212 \pi^2 T^2 - 167) \bigg] \,.
\end{split}
\end{equation}
At the fourth order ($n=4$), the explicit expressions expand to:
\begin{equation}
\begin{split}
    E^{(a)}_4(\lambda) &= \frac{4 \big(5 - 52 a^2 + 7 a^4 - 64 i a \pi T - 64 i a^3 \pi T \big)}{1+a^2} \,, \\[6pt]
    E^{(a)}_4(\lambda^2) &= \frac{4}{(1+a^2)^2} \bigg[ 27 + 43 a^8 - 960 i a \pi T + 7744 i a^3 \pi T + 7360 i a^5 \pi T \\
    & \quad - 1344 i a^7 \pi T + a^4 (3546 - 11008 \pi^2 T^2) - 128 a^2 (6 + 43 \pi^2 T^2) \\
    & \quad - 32 a^6 (35 + 172 \pi^2 T^2) \bigg] \,,
\end{split}
\end{equation}
\begin{equation}
\begin{split}
    E^{(a)}_4(\lambda^3) &= \frac{16}{(1+a^2)^3} \bigg[ 9 + 15 a^{12} - 864 i a \pi T - 1376 i a^{11} \pi T - 72 a^2 (9 + 184 \pi^2 T^2) \\
    & \quad + 32 i a^3 \pi T (609 + 1408 \pi^2 T^2) + 32 i a^9 \pi T (905 + 1408 \pi^2 T^2) \\
    & \quad + 64 i a^5 \pi T (2112 \pi^2 T^2 - 967) + 64 i a^7 \pi T (2112 \pi^2 T^2 - 811) \\
    & \quad - 8 a^{10} (145 + 2344 \pi^2 T^2) + 19 a^8 (649 + 3456 \pi^2 T^2) \\
    & \quad + 16 a^6 (10896 \pi^2 T^2 - 1439) + a^4 (7869 + 76672 \pi^2 T^2) \bigg] \,,
\end{split}
\end{equation}
\begin{equation}
\begin{split}
    E^{(a)}_4(\lambda^4) &= \frac{512 a}{(1+a^2)^4} \bigg[ -18 i \pi T - 30 i a^{14} \pi T - 9 a (1 + 76 \pi^2 T^2) - a^{13} (25 + 1116 \pi^2 T^2) \\
    & \quad + 6 i a^{12} \pi T (295 + 1472 \pi^2 T^2) + 2 i a^2 \pi T (459 + 3008 \pi^2 T^2) \\
    & \quad - 14 i a^8 \pi T (6016 \pi^2 T^2 - 651) - 2 i a^{10} \pi T (7191 + 6208 \pi^2 T^2) \\
    & \quad - 2 i a^4 \pi T (4281 + 10432 \pi^2 T^2) - 6 i a^6 \pi T (14976 \pi^2 T^2 - 2633) \\
    & \quad + 6 a^3 (51 + 1940 \pi^2 T^2 + 2240 \pi^4 T^4) + 2 a^{11} (361 + 9180 \pi^2 T^2 + 6720 \pi^4 T^4) \\
    & \quad + a^5 (53760 \pi^4 T^4 - 19908 \pi^2 T^2 - 2375) + a^9 (53760 \pi^4 T^4 - 5172 \pi^2 T^2 - 4311) \\
    & \quad + a^7 (80640 \pi^4 T^4 - 56880 \pi^2 T^2 + 5692) \bigg] \,.
\end{split}
\end{equation}
For brevity, we omit the fully expanded $n \ge5$ terms of the angular sector, as they share the identical algebraic structure and can be systematically derived utilizing the Vieta's formulas defined in the main text.

\section{The linearity of \texorpdfstring{$f_{1n}$, $f_{2n}$ and $f_{3n}$}{f1n, f2n and f3n}}\label{app_linear_f1f2f3}

\subsection{Preliminaries and Notation}
In this appendix, we present several proofs that are essential for the derivations in section~\ref{3Drotating}. For convenience and self-consistency, we briefly restate the relevant expressions and notation introduced in the main text.

Consider three-dimensional stationary and axisymmetric rotating black holes described by the metric \eqref{3rBH}, coupled to a free scalar field $\Phi$ of mass $m$. The radial part of its equation of motion takes the form

\begin{equation}\label{equ_3D_rotating_KGequ}
\begin{split}
    f_1(z) f_2(z) \psi''(z)
+ \left( \frac{1}{2} f_1(z) f_2'(z) + \frac{1}{2} f_2(z) f_1'(z) - \frac{f_1(z) f_2(z)}{z} \right) \psi'(z)&\\[6pt]
+ \left( (\omega - k f_3(z))^2 - f_1(z)\frac{k^2 z^2 + M^2}{z^2} \right) \psi(z)& = 0,
\end{split}
\end{equation}
where $\Phi = e^{-i\omega t + i k x}\psi$. For later convenience, we have multiplied the original radial equation \eqref{EOM2} by an overall factor $f_1(z)f_2(z)$.

The metric functions $f_1(z)$, $f_2(z)$, and $f_3(z)$ are analytic near the horizon, which we fix at $z=1$. Their Taylor expansions take the form
\begin{equation}\label{equ_f1f2f3_nearhorizon_expansion}
f_1(z) = \sum_{n=1}^{\infty} f_{1n}(z-1)^n, \qquad
f_2(z) = \sum_{n=1}^{\infty} f_{2n}(z-1)^n, \qquad
f_3(z) = \Omega_H + \sum_{n=1}^{\infty} f_{3n}(z-1)^n,
\end{equation}
where $f_{30}=\Omega_H$ is identified as the angular velocity of the horizon. The associated Hawking temperature $T$ is determined by the relation
\begin{equation}
f_{11} = \frac{(4\pi T)^2}{f_{21}}.
\end{equation}
We further define $\mathfrak{w} \equiv \omega - k \Omega_H$.

We now consider the ingoing Frobenius series solution for $\psi(z)$ near the horizon,
\begin{equation}\label{equ_series_ingoing_solution}
\psi(z) = (z-1)^{\rho} \sum_{j=0}^{n} \phi_j (z-1)^j,
\end{equation}
where $\rho=-i\mathfrak{w}/(4\pi T)$.

Substituting the near-horizon expansions \eqref{equ_f1f2f3_nearhorizon_expansion} and the ansatz \eqref{equ_series_ingoing_solution} into Eq.~\eqref{equ_3D_rotating_KGequ}, and collecting terms at each order $(z-1)^j$ for $j\in\{1,2,\ldots,n\}$, yields a linear system for the coefficients $\phi_j$. This system can be written compactly as
\begin{equation}
\mathbb{M}\cdot \phi^{(n)} = 0,
\end{equation}
where $\mathbb{M}$ is an $n\times(n+1)$ matrix.\footnote{Written explicitly row by row, this system coincides with Eq.~\eqref{S2} in the main text.} We denote by $M_{ij}$ the entry in the $i^{\text{th}}$ row and $j^{\text{th}}$ column of $\mathbb{M}$, and by $\mathcal{M}^{(n)}$ the first $n\times n$ submatrix of $\mathbb{M}$.

For later convenience, we introduce the following coefficient functions appearing in Eq. \eqref{equ_3D_rotating_KGequ}:
\begin{equation}
\begin{aligned}\label{equ_coe_ABC}
A(z) &= f_1(z)\,f_2(z)
     = \sum_{m=2}^{\infty} A_m (z-1)^m, \\[0.5em]
B(z) &= \frac{1}{2}\bigl(f_1(z)\,f_2(z)\bigr)'
       - \frac{f_1(z)\,f_2(z)}{z}
     = \sum_{m=1}^{\infty} B_m (z-1)^m, \\[0.5em]
C(z) &= \bigl(\omega - k f_3(z)\bigr)^2
       - f_1(z)\,\frac{k^2 z^2 + m^2}{z^2}
     = \sum_{m=0}^{\infty} C_m (z-1)^m .
\end{aligned}
\end{equation}

With these definitions, the matrix element $M_{KS}$, corresponding to the coefficient of $(z-1)^{\rho+K}\psi^{S-1}$, can be written in terms of $A_m$, $B_m$, and $C_m$ as
\begin{equation}\label{equ_general_formula_Mij}
M_{KS}
= A_{K-S+3}(\rho+S-1)(\rho+S-2)
+ B_{K-S+2}(\rho+S-1)
+ C_{K-S+1}.
\end{equation}
Specializing to $K=n$ and $S=n+1$, this expression reduces to
\begin{equation}
M_{n\,n+1}
= A_2(\rho+n)(\rho+n-1)
+ B_1(\rho+n)
+ C_0 .
\end{equation}
Substituting the explicit expressions for $A_2$, $B_1$, $C_0$, and the ingoing exponent $\rho$, one finds
\begin{equation}
M_{n\,n+1}
= 8\pi T\,n \bigl(2\pi T\,n - i\mathfrak{w}\bigr).
\end{equation}

It follows immediately that, at the pole-skipping frequency $\mathfrak{w}=\mathfrak{w}_n=-i\,2\pi Tn$, one has $M_{n\,n+1}=0$. Since this is the only nonvanishing entry in the last column, the entire last column of the matrix $\mathbb{M}$ vanishes. Consequently, the original linear system $\mathbb{M}\cdot\phi^{(n)}=0$ reduces to
\begin{equation}
\mathcal{M}^{(n)}\,\phi'=0,
\end{equation}
where $\phi'=(\phi_0,\ldots,\phi_{n-1})^T$.

The pole-skipping momentum is then determined by the condition
\begin{equation}
\det\bigl(\mathcal{M}^{(n)}\bigr)=0,
\end{equation}
which can be regarded as a polynomial equation in $k$ of degree $2n$. To see this, note that, according to Eq.~\eqref{equ_coe_ABC}, the dependence on $k$ enters exclusively through the coefficients $C_m$, each of which is a polynomial in $k$ of degree at most $2$. The determinant can be expanded as a sum of products of matrix elements. Among these terms, the product of all diagonal elements,
\begin{equation}
\prod_{K=1}^{n} M_{KK},
\end{equation}
contains one factor of $C_1$ from each $M_{KK}$ and therefore yields a polynomial in $k$ of degree $2n$. The coefficient of $k^{2n}$ arises from the product of the $k^2$ terms in each $C_1$.\footnote{At the pole-skipping frequency, $C_1=-f_{11}k^2-2k\mathfrak{w}_n f_{31}-f_{11}m^2$.} All other contributions to the determinant necessarily involve at least one off-diagonal element replacing a diagonal one, and hence give terms of strictly lower degree in $k$. It follows that $\det\bigl(\mathcal{M}^{(n)}\bigr)$ is a polynomial $P_n(k^{2n})$ of degree exactly $2n$.

We can further express the determinant $\det\bigl(\mathcal{M}^{(n)}\bigr)=P_n(k^{2n})$ explicitly as
\begin{equation}\label{equ_Pnk}
P_n(k^{2n})\equiv V_{n,2n}k^{2n}+V_{n,2n-1}k^{2n-1}+\cdots+V_{n,1}k+V_{n,0},
\end{equation}
where each coefficient $V_{n,m}$ depends on $f_{1i}$, $f_{2i}$, and $f_{3i}$. The roots of this polynomial are denoted by $k_{n,1},\ldots,k_{n,2n}$. Applying Vieta’s formulas and defining $v_{n,m}=(-1)^{\,n-m}V_{n,m}$, we obtain
\begin{equation}\label{equ_vieta_equation}
\mathcal{M}^{(n)}_{m}\equiv E_n(k^m)=\frac{v_{n,2n-m}}{v_{n,2n}},
\end{equation}
for $m=1,2,\ldots,2n$, where $E_n(k^m)$ denotes the elementary symmetric polynomial of degree $m$ in the roots $k_{n,1},\ldots,k_{n,2n}$, namely
\begin{equation}
E_n(k^m)
=\sum_{1\le i_1<i_2<\cdots<i_m\le 2n}
k_{n,i_1}\,k_{n,i_2}\cdots k_{n,i_m}.
\end{equation}

In what follows, we establish the following result.

\begin{theorem}\label{theo_linear_f123}
For $n>1$, the equations $\mathcal{M}^{(n)}_{2n}$, $\mathcal{M}^{(n)}_{2n-1}$, and $\mathcal{M}^{(n)}_{2n-2}$ are respectively linear in the \textbf{unknowns} $(f_{1n},f_{2n})$, in $(f_{1n},f_{2n},f_{3n})$, and again in $(f_{1n},f_{2n})$.
\end{theorem}

Throughout the analysis, all lower-order coefficients $f_{1i}$, $f_{2i}$, and $f_{3i}$ with $i<n$ are treated as fixed parameters.

\subsection{Proof}

To prove \textbf{Theorem 1}, we first show that $P_n(k^{2n})$ is linear in the variables $(f_{1n},f_{2n},f_{3n})$.

We begin by recalling from Eq.~\eqref{equ_coe_ABC} that the coefficients $A_{m+1}$, $B_m$, and $C_m$ are linear in $(f_{1m},f_{2m})$, in $(f_{1m},f_{2m})$, and in $(f_{1m},f_{3m})$, respectively.

To establish the linearity explicitly, we first examine the matrix element $M_{n1}$, whose explicit form follows from Eq.~\eqref{equ_coe_ABC}:
\begin{equation}\label{equ_Mn1}
M_{n1}=A_{n+2}\rho(\rho-1)+B_{n+1}\rho+C_n.
\end{equation}
From the definition of $B(x)$ in Eq.~\eqref{equ_coe_ABC}, the coefficient $B_{n+1}$ can be written as
\begin{equation}\label{equ_relation_AB}
B_{n+1}=\frac{n+2}{2}A_{n+2}+\cdots,
\end{equation}
where the ellipsis denotes terms that depend only on $A_m$ with $m\le n+1$, and hence only on $f_{1i}$ and $f_{2i}$ with $i\le n$. Substituting this expression into Eq.~\eqref{equ_Mn1}, the coefficient multiplying $A_{n+2}$ becomes $\rho(\rho+n/2)$.

Naively, $A_{n+2}$ depends on $f_{1\,n+1}$ and $f_{2\,n+1}$. However, at the pole-skipping frequency $\mathfrak{w}_n=-i\,2\pi T n$, corresponding to $\rho=-n/2$, this coefficient vanishes identically. As a result, the dependence of $M_{n1}$ on $f_{1\,n+1}$ and $f_{2\,n+1}$ drops out. The remaining contributions to $M_{n1}$ arise from $C_n$, which involves $(f_{1n},f_{3n})$, and from $A_{n+1}$, which involves $(f_{1n},f_{2n})$. Hence, $M_{n1}$ is linear in the unknowns $f_{1n}$, $f_{2n}$, and $f_{3n}$.

There is no other matrix element that is linear in $f_{3n}$. Indeed, the condition $K-S+1\ge n$ admits only the solution $(K=n,S=1)$. The remaining entries that are linear in $f_{1n}$ and $f_{2n}$ are $M_{n2}$ and $M_{n-1\,1}$, as can be directly verified from Eq.~\eqref{equ_coe_ABC}.

We now expand the determinant along the last row:
\begin{equation}\label{equ_cofactor_expansion}
P_n(k^{2n})=M_{n1}\Delta_{n1}+M_{n2}\Delta_{n2}+\sum_{j=3}^n M_{nj}\Delta_{nj}.
\end{equation}
In the first term, $M_{n1}\Delta_{n1}$, the matrix element $M_{n1}$ is linear in $f_{1n}$, $f_{2n}$, and $f_{3n}$, while the associated cofactor $\Delta_{n1}$ takes the explicit form
\begin{equation}\label{equ_Deltan1}
\Delta_{n1}=(-1)^{n+1}\prod_{S=1}^{n-1} 8\pi T\,S \bigl(2\pi T\,S - i\mathfrak{w}_n\bigr),
\end{equation}
which is independent of the unknown coefficients and therefore constant. Consequently, $M_{n1}\Delta_{n1}$ is linear in $f_{1n}$, $f_{2n}$, and $f_{3n}$.

In the second term, $M_{n2}\Delta_{n2}$, the matrix element $M_{n2}$ (for $n>2$) is linear in $f_{1n}$ and $f_{2n}$.\footnote{For $n=2$, one has $(\frac{-i\mathfrak{w}}{4\pi T}+2-1)\big|_{\mathfrak{w}=-4i\pi T}=0$, so that $M_{22}=C_1$.} The corresponding cofactor is given by
\begin{equation}\label{equ_Deltan2}
\Delta_{n2}=(-1)^{n+2}M_{11}\prod_{S=1}^{n-2} 8\pi T\,S \bigl(2\pi T\,S - i\mathfrak{w}_n\bigr),
\end{equation}
which is likewise constant. Consequently, the full term $M_{n2}\Delta_{n2}$ is linear in the variables $f_{1n}$ and $f_{2n}$.

For the remaining terms $M_{ni}\Delta_{ni}$ with $i>2$, the entries $M_{ni}$ do not depend on $f_{1n}$, $f_{2n}$, or $f_{3n}$. Although each $\Delta_{ni}$ contains $M_{n-1\,1}$ in its first column and therefore appears to be linear in $f_{1n}$ and $f_{2n}$ as well, a direct expansion of $\Delta_{ni}$ along this column shows that the cofactor multiplying $M_{n-1\,1}$ is the determinant of a matrix whose last column vanishes identically. Consequently, for $2<i<n$, this contribution is zero, and these terms do not involve the unknown coefficients.

A different situation arises for $\Delta_{nn}$. In this case, the cofactor associated with $M_{n-1\,1}$ does not vanish; instead, it takes the explicit form
\begin{equation}\label{equ_Delta_nn}
\prod_{S=1}^{n-2} 8\pi T\,S \bigl(2\pi T\,S - i\mathfrak{w}_n\bigr).
\end{equation}
As a result, the last term in the expansion, $M_{nn}\Delta_{nn}$, is again linear in $f_{1n}$ and $f_{2n}$.

Collecting all contributions, we conclude that $P_n(k^{2n})$ is linear in the coefficients $f_{1n}$, $f_{2n}$, and $f_{3n}$. Furthermore, when the determinant is expanded along the last row, this linear dependence arises entirely from the entries $M_{n1}$, $M_{n2}$, and the cofactor $\Delta_{nn}$.

Our next step is to show that the Vieta equations defined in Eq.~\eqref{equ_vieta_equation} of degrees $2n$, $2n-1$, and $2n-2$ are linear in the unknowns specified in \textbf{Theorem 1}. Since $v_{n,2n}=f_{11}^{n}$, it suffices to prove that the coefficients of the $k^{0}$ (constant), $k^{1}$, and $k^{2}$ terms in $P_n(k^{2n})$, namely $v_{n,0}$, $v_{n,1}$, and $v_{n,2}$, are linear in $(f_{1n},f_{2n})$, $(f_{1n},f_{2n},f_{3n})$, and $(f_{1n},f_{2n})$, respectively.

According to the analysis in the previous section, the only contributions to $P_n(k^{2n})$ that are linear in $f_{1n}$, $f_{2n}$, or $f_{3n}$ arise from three terms in the cofactor expansion \eqref{equ_cofactor_expansion}, namely
$M_{n1}\Delta_{n1}$, $M_{n2}\Delta_{n2}$, and $M_{nn}\Delta_{nn}$. Moreover, the dependence on $k$ and $k^{2}$ originates exclusively from the coefficients $C_m$. We now analyze these three contributions in turn.

For the term $M_{n1}\Delta_{n1}$, the linear dependence on the unknowns comes entirely from $M_{n1}$, whose explicit form is given in Eq.~\eqref{equ_Mn1}. Examining its $k$ dependence, we find the following contributions:
\begin{itemize}
\item From $C_n$, the coefficient $f_{3n}$ appears linearly in the $k$ term through $-2\mathfrak{w} k f_{3n}$, while $f_{1n}$ appears linearly in both the $k^{2}$ and $k^0$ terms through $-f_{1n}(k^{2}+m^{2})$.
\item From the combined contributions of $A_{n+2}$ and $B_{n+1}$, using the relations derived around Eq.~\eqref{equ_relation_AB}, both $f_{1n}$ and $f_{2n}$ appear linearly in the $k^0$-term.
\end{itemize}

Next, consider the term $M_{n2}\Delta_{n2}$ for $n>2$. Here the linear dependence on the unknowns arises solely from $M_{n2}$, which takes the form
\begin{equation}\label{equ_Mn2}
M_{n2}=A_{n+1}(\rho+1)\rho+B_n(\rho+1)+C_{n-1}.
\end{equation}
In this expression, $C_{n-1}$ does not contribute to the dependence on the unknown coefficients. Examining the terms $A_{n+1}$ and $B_n$, we find that both $f_{1n}$ and $f_{2n}$ enter linearly. Moreover, through the multiplicative factor $M_{11}$ within the expression for $\Delta_{n2}$ \eqref{equ_Deltan2}, this linear dependence is distributed among the $k^{0}$, $k^{1}$, and $k^{2}$ terms.

Finally, we examine the contribution from $M_{nn}\Delta_{nn}$. In this case, the linear dependence on the unknowns originates from the factor $M_{n-1\,1}$, which appears in the cofactor expansion of $\Delta_{nn}$ and is given by
\begin{equation}\label{equ_Mnminus11}
M_{n-1\,1}=A_{n+1}\rho(\rho-1)+B_n\rho+C_{n-1}.
\end{equation}
The relevant contribution to $M_{nn}\Delta_{nn}$ is of the form $M_{nn}M_{n-1\,1}\cdots$, where the ellipsis denotes factors that are independent of $k$ and of the unknown coefficients. The entry $M_{n-1\,1}$ contains $A_{n+1}$ and $B_n$, implying that $f_{1n}$ and $f_{2n}$ appear linearly in the $k^0$-term. Meanwhile, $M_{nn}$ contains $C_1$, which contributes both $k$ and $k^{2}$ terms. Their product therefore yields linear dependence on $f_{1n}$ and $f_{2n}$ in the $k^0$, $k$, and $k^{2}$ terms.

Combining all contributions, we conclude that the coefficients of the $k^{0}$, $k^{1}$, and $k^{2}$ terms in $P_n(k^{2n})$, namely $v_{n,0}$, $v_{n,1}$, and $v_{n,2}$, are linear in $(f_{1n},f_{2n})$, $(f_{1n},f_{2n},f_{3n})$, and $(f_{1n},f_{2n})$, respectively. This establishes the desired linearity of the corresponding Vieta equations and completes the proof of \textbf{Theorem 1}.

\section{Explicit expressions of third-order derivatives of three-dimensional rotating  metric}\label{f1233}
In section~\ref{sec:iter_rotating}, we iteratively derived the explicit expressions for the metric derivatives up to the second order. We now proceed to determine the third-order terms, which are governed by the system \eqref{V2}
\begin{equation}
    \begin{split}
        V_{3,0}(f_{a3})=V_{3,6}\,E_3(k^6)\,,\quad
        V_{3,1}(f_{a3})=-V_{3,6}\,E_3(k^5)\,,\quad
        V_{3,2}(f_{a3})=V_{3,6}\,E_3(k^4)\,.
    \end{split}
\end{equation}
By substituting the lower-order results \eqref{f131} and \eqref{f1232} into this system, we obtain the linear equations for $f_{a3}$, whose solutions can be explicitly expressed as
\begin{equation}
    \begin{split}
        f_{13}=&\frac{\mathfrak{w}_1^2}{24(E_1(k^2)-m^2)^5}\bigg[ 1242 E_1(k)^2 E_1(k^2)^3 + 540 E_1(k^2)^4 - 216 E_1(k^2)^3 E_2(k^2)\\[4pt]
        &\quad- 252 E_1(k) E_1(k^2)^2 E_2(k^3) + 
 108 E_1(k)^2 E_1(k^2) E_2(k^4) + 36 E_1(k^2)^2 E_2(k^4) \\[4pt]
 &\quad- 36 E_1(k^2) E_2(k^2) E_2(k^4) + 9 E_2(k^4)^2 + 
 18 E_1(k^2)^2 E_3(k^4) - 
 2 E_1(k^2) E_3(k^6) \\[4pt]
 &\quad+ \Big(324 E_1(k)^4 E_1(k^2) - 1998 E_1(k)^2 E_1(k^2)^2 - 1020 E_1(k^2)^3 - 292 E_1(k^2)^2 E_2(k^2) \\[4pt]
 &\quad+216 E_1(k)^2 E_1(k^2) E_2(k^2) +  36 E_1(k^2) E_2(k^2)^2 + 
    476 E_1(k) E_1(k^2) E_2(k^3)\\[4pt]
& \quad- 54 E_1(k)^2 E_2(k^4) + 64 E_1(k^2) E_2(k^4) + 18 E_2(k^2) E_2(k^4) - 
    34 E_1(k^2) E_3(k^4) \\[4pt]
&\quad+ 2 E_3(k^6)\Big)m^2 + \Big(-243 E_1(k)^4 + 876 E_1(k)^2 E_1(k^2) + 
    564 E_1(k^2)^2 - 27 E_2(k^2)^2 \\[4pt]
    &\quad+ 162 E_1(k)^2 E_2(k^2) - 116 E_1(k^2) E_2(k^2)  - 
    224 E_1(k) E_2(k^3) - 82 E_2(k^4) \\[4pt]
    &\quad+ 16 E_3(k^4)\Big) m^4 + \Big(-66 E_1(k)^2 - 80 E_1(k^2) + 
    22 E_2(k^2)\Big) m^6 + 5 m^8
        \bigg],
    \end{split}
\end{equation}
\begin{equation}
    \begin{split}
        f_{23}=&\frac{1}{24(E_1(k^2)-m^2)^3}\bigg[ 1242 E_1(k)^2 E_1(k^2)^3 + 540 E_1(k^2)^4 - 216 E_1(k^2)^3 E_2(k^2) \\[4pt]
        &\quad- 252 E_1(k) E_1(k^2)^2 E_2(k^3) + 
 108 E_1(k)^2 E_1(k^2) E_2(k^4) + 36 E_1(k^2)^2 E_2(k^4) \\[4pt]
 &\quad- 36 E_1(k^2) E_2(k^2) E_2(k^4) + 9 E_2(k^4)^2 + 
 18 E_1(k^2)^2 E_3(k^4) - 
 2 E_1(k^2) E_3(k^6) \\[4pt]
 &\quad+ \Big(324 E_1(k)^4 E_1(k^2) - 1998 E_1(k)^2 E_1(k^2)^2  - 
    216 E_1(k)^2 E_1(k^2) E_2(k^2) \\[4pt]
    &\quad- 1020 E_1(k^2)^3+ 292 E_1(k^2)^2 E_2(k^2) + 36 E_1(k^2) E_2(k^2)^2 - 54 E_1(k)^2 E_2(k^4)\\[4pt]
    &\quad+ 
    476 E_1(k) E_1(k^2) E_2(k^3)  + 64 E_1(k^2) E_2(k^4) + 18 E_2(k^2) E_2(k^4) + 2 E_3(k^6)\\[4pt]
    &\quad- 
    34 E_1(k^2) E_3(k^4) \Big) m^2 + \Big(-243 E_1(k)^4 + 876 E_1(k)^2 E_1(k^2) + 
    564 E_1(k^2)^2 \\[4pt]
    &\quad+ 162 E_1(k)^2 E_2(k^2) - 116 E_1(k^2) E_2(k^2) - 27 E_2(k^2)^2 - 
    224 E_1(k) E_2(k^3)\\[4pt]
    &\quad- 82 E_2(k^4) + 16 E_3(k^4)\Big) m^4 + \Big(-66 E_1(k)^2 - 80 E_1(k^2) + 
    22 E_2(k^2)\Big) m^6 + 5 m^8
        \bigg],
    \end{split}
\end{equation}
\begin{equation}
    \begin{split}
        f_{33}=&\frac{\mathfrak{w}_1}{48(E_1(k^2)-m^2)^5}\bigg[540 E_1(k) E_1(k^2)^4 - 108 E_1(k^2)^3 E_2(k^3) + 36 E_1(k) E_1(k^2)^2 E_2(k^4)\\[4pt]
        &\qquad\quad - 18 E_1(k^2) E_2(k^3) E_2(k^4) + 
 9 E_1(k) E_2(k^4)^2  - 
 2 E_1(k) E_1(k^2) E_3(k^6) \\[4pt]
 &\qquad\quad+ 6 E_1(k^2)^2 E_3(k^5)+ \Big(486 E_1(k)^3 E_1(k^2)^2 - 
    140 E_1(k) E_1(k^2)^2 E_2(k^2)\\[4pt]
    &\qquad\quad- 1020 E_1(k) E_1(k^2)^3  - 82 E_1(k)^2 E_1(k^2) E_2(k^3) + 216 E_1(k^2)^2 E_2(k^3) \\[4pt]
    &\qquad\quad+ 
    18 E_1(k^2) E_2(k^2) E_2(k^3) + 54 E_1(k)^3 E_2(k^4) + 64 E_1(k) E_1(k^2) E_2(k^4) \\[4pt]
    &\qquad\quad- 18 E_1(k) E_2(k^2) E_2(k^4) + 
    18 E_2(k^3) E_2(k^4) + 2 E_1(k) E_1(k^2) E_3(k^4)\\[4pt] 
    &\qquad\quad- 12 E_1(k^2) E_3(k^5) + 
    2 E_1(k) E_3(k^6)\Big) m^2 + \Big(81 E_1(k)^5 - 474 E_1(k)^3 E_1(k^2) \\[4pt]
    &\qquad\quad+ 564 E_1(k) E_1(k^2)^2 - 
    54 E_1(k)^3 E_2(k^2) + 136 E_1(k) E_1(k^2) E_2(k^2) \\[4pt]
    &\qquad\quad+ 9 E_1(k) E_2(k^2)^2 + 82 E_1(k)^2 E_2(k^3) - 
    126 E_1(k^2) E_2(k^3) - 18 E_2(k^2) E_2(k^3)\\[4pt]
    &\qquad\quad- 82 E_1(k) E_2(k^4) - 2 E_1(k) E_3(k^4) + 
    6 E_3(k^5)\Big) m^4 + \Big(42 E_1(k)^3 \\[4pt]
    &\qquad\quad- 80 E_1(k) E_1(k^2) - 14 E_1(k) E_2(k^2) + 
    18 E_2(k^3)\Big) m^6 + 5 E_1(k) m^8
        \bigg].
    \end{split}
\end{equation}

\bibliographystyle{JHEP}
\bibliography{references}

\end{document}